\documentclass[preprint]{aastex}

\newcommand{\kms}{km~s$^{-1}$}

\newcommand{\subsun}{\mbox{$_{\odot}$}}
\newcommand{\teff}{$T_{eff}$}
\newcommand{\grav}{log($g$)}
\newcommand{\etal}{{\it et al.\/}}
\newcommand{\eqw}{$W_{\lambda}$}

\newcommand{\ourglob}{NGC~7492}

\begin{document}

\title{Outer Versus Inner Halo Globular Clusters: \ourglob\
Abundances\altaffilmark{1}}

\author{Judith G. Cohen\altaffilmark{2} and Jorge Melendez\altaffilmark{2}}

\altaffiltext{1}{Based in part on observations obtained at the
W.M. Keck Observatory, which is operated jointly by the California 
Institute of Technology, the University of California, and the
National Aeronautics and Space Administration.}

\altaffiltext{2}{Palomar Observatory, Mail Stop 105-24,
California Institute of Technology, Pasadena, Ca., 91125, jlc(jorge)@astro.caltech.edu}

\begin{abstract}

We have carried out a detailed abundance analysis for 21 elements 
in a sample of four RGB stars in the outer halo globular cluster
\ourglob\ ($R_{GC}$ 25 kpc); we find [Fe/H] = $-1.82$ dex inferred from Fe~I 
lines ($-1.79$ from Fe~II) using high dispersion 
(R=$\lambda / \Delta \lambda$=35,000) spectra obtained
with HIRES at the Keck Observatory.
Most elements show no sign of star-to-star
variation within our limited sample.
We have, however, detected an anti-correlation between 
O and Na abundances
similar to that seen in our previous analyses of inner halo
GCs as well as in studies of relatively nearby GCs by others.

We compare the abundance ratios in \ourglob\ with those we
previously determined for the much closer old halo GCs M3 and M13. 
After making corrections
for trends of abundance ratio with metallicity
characteristic of halo stars,  we find that for
for these three GCs for
each of the elements in common we deduce identical
abundance ratios with respect to Fe to within the probable measurement
uncertainties. Thus,
the chemical history of the outer halo as exemplified by
the metal-poor outer halo globular cluster \ourglob\
is indistinguishable from that of the inner halo,
exemplified by M3 and M13, at least through
the epoch of formation of these old globular clusters.
This applies to the neutron capture
processes as well.

\end{abstract}


\section{Introduction}

Abundance determinations of stars in Galactic globular clusters can provide 
valuable information about important astrophysical processes such as
stellar evolution, stellar structure, Galactic chemical evolution and
the formation of the Milky Way.  With the advent of efficient high
resolution spectrographs on 8-10m telescopes, it is now possible
to reach at least the luminous RGB stars in 
even the most distant Galactic globular clusters (GCs).
In our previous work in this area
we have explored the abundances for large samples of stars
in the canonical relatively
nearby GCs M71 \citep{ramirez01,ramirez02}, M5 \citep{ramirez03}
M3, and M13 \citep{cohen05}, as well as in
Pal~12 \citep{cohen04}, a cluster associated with the Sgr dwarf
galaxy \citep{irwin99,dinescu00,ibata01}.  For the
case of Pal~12, in addition to the
difference in age of several Gyr, it being younger than
the bulk of the Galactic GCs \citep{rosenberg98,rosenberg99}, we found
evidence for substantive differences between the chemical history
of Pal~12 and that of the ``normal'' halo GCs of similar 
[Fe/H]\footnote{The 
standard nomenclature is adopted; the abundance of
element $X$ is given by $\epsilon(X) = N(X)/N(H)$ on a scale where
$N(H) = 10^{12}$ H atoms.  Then
[X/H] = log[N(X)/N(H)] $-$ log[N(X)/N(H)]\subsun, and similarly
for [X/Fe].}.

In the present work, we study the outer halo GC \ourglob, a cluster which
is not suspected (yet) of being part of any known stream or otherwise abnormal.
We compare the abundance ratios deduced from high resolution, high
signal-to-noise ratio for four
giants in this distant outer halo GC with those from our
recent analysis of a large sample of stars in the relatively nearby
and well studied GCs M3 and M13, which have metallicities close to
that of \ourglob.  We look for evidence in the
deduced abundance ratios of some difference in the formation mechanisms
or chemical history of GCs in the ``normal'' outer halo.

There has been no
previous high dispersion study of \ourglob.  \cite{zinn84}
obtained [Fe/H] = $-1.5\pm0.3$ dex using their narrow band Q39 photometric system,
\cite{smith84} derived $-1.34\pm0.25$ dex from the ${\Delta}S$
method applied to two RR Lyrae variables in the cluster, while
moderate dispersion spectroscopy in the region of the infrared Ca triplet
by \cite{rutledge97b} gave [Fe/H] = $-1.70\pm0.06$ dex.

\section{Stellar Sample, Observations and \teff\ Determination}

\begin{figure}
\epsscale{0.9}
\figurenum{1}
\plotone{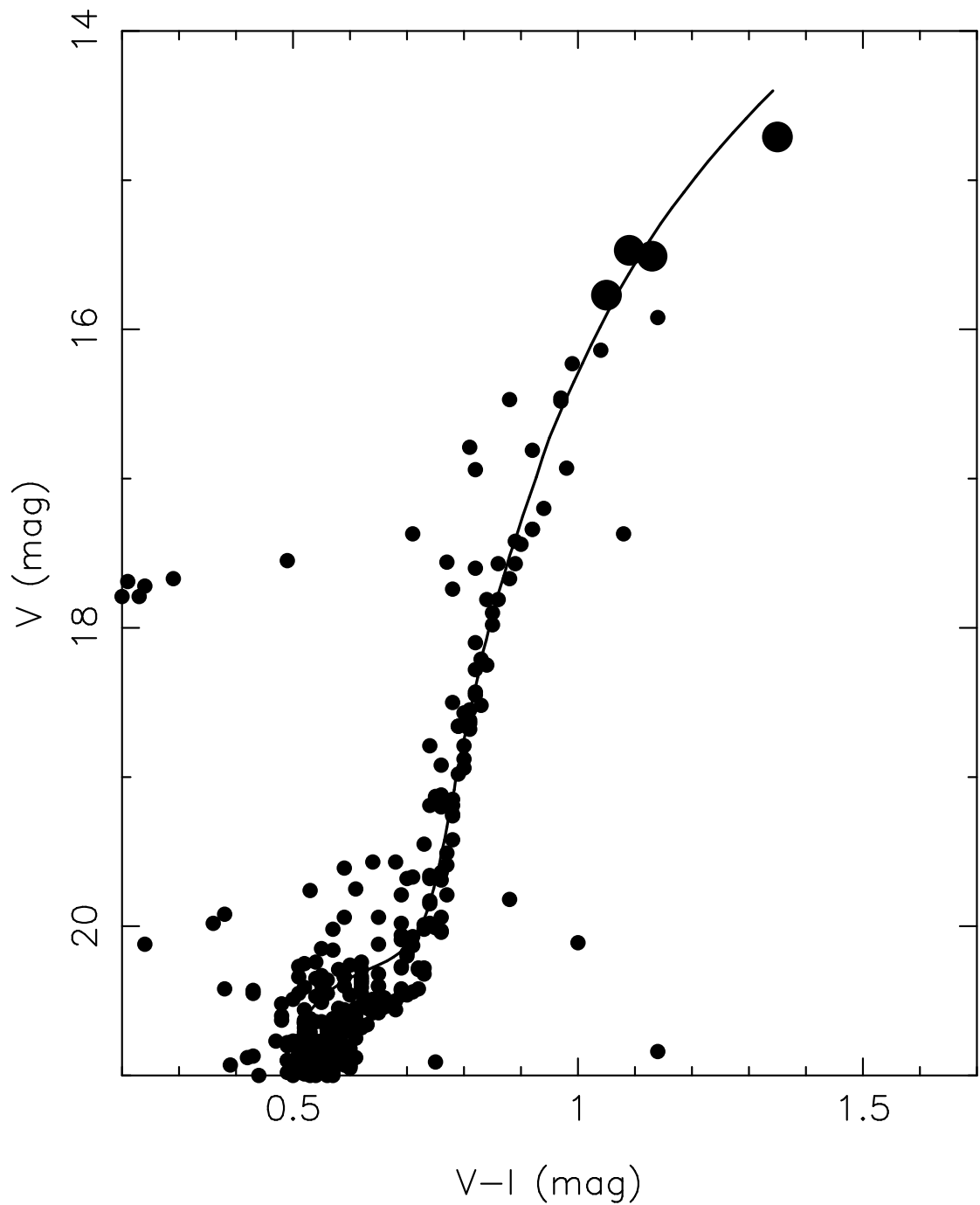}
\caption[]{The  $V-I$ CMD of \ourglob: the four RGB stars observed
with HIRES (big filled circles) are shown superposed on 
a 12 Gyr isochrone from \cite{yi01} with
[Fe/H] $-1.7$ dex 
shifted to the distance of \ourglob.  
The small circles denote stars from \cite{buonanno87} in this GC roughly
transformed from $B,V$ to $V,I$. 
\label{figure_cmd}}
\end{figure}

Given the large distance of \ourglob, we select the brightest
possible stars on the upper RGB for observation, making no
attempt to reach fainter luminosities.  These stars were picked
from the photometric study of this cluster by
\cite{cuffey61}.  Since this
is a rather sparse cluster, these stars, which are the four
brightest suspected members, lie distributed along
the upper RGB, not concentrated at the RGB tip.
The positions of these stars in a $V,I$ CMD
are illustrated in Figure~\ref{figure_cmd}
superposed on the predicted cluster isochrone from the \cite{yi01} evolutionary
tracks; see Figure~10 of \cite{buonanno87} for a $B,V$ CMD of \ourglob.
Although the cluster is distant, there are so few luminous
giants that they must be observed as single stars; two
of them cannot be fit into a single 7 arcsec long slit.
Spectra were obtained with HIRES \citep{vogt94} at 
the Keck Observatory 20-22 August 2003.    The
instrument configuration 
covered the range 4650 to 7010~\AA,  with small gaps
between the orders at the red end.  This is the ``yellow''
configuration described in \cite{cohen05}.
These data were reduced using a combination of Figaro 
\citep{shortridge93} scripts and
the software package MAKEE \footnote{MAKEE was developed
by T.A. Barlow specifically for reduction of Keck HIRES data.  It is
freely available on the world wide web at the
Keck Observatory home page, http://www2.keck.hawaii.edu:3636/.}.

The desired minimum SNR was 90 over a 4 pixel resolution element for
a wavelength near the center of the HIRES detector.
This is calculated strictly from the counts in the object spectrum, and 
excludes noise from cosmic ray hits, sky subtraction, flattening problems, etc. 
Since the nights were relatively dark, sky subtraction is not an issue except at
the specific wavelengths corresponding to strong night sky emission lines, 
such as the Na D doublet. 
The seeing was extremely
good during these nights, making the exposures shorter than normal, and
enabling us to reach this goal for the stars reported here.
Table~\ref{table_sample} gives details of the HIRES exposures for each star, 
with the total exposure time for each object.
All long integrations were broken 
up into separate exposures, each 1200 sec long, to optimize cosmic
ray removal.
The last column of the table gives
the heliocentric radial velocity for each star,
measured from the HIRES spectra; see \cite{ramirez03} for
the details of the procedure used to determine $v_r$.

The  radial velocity of \ourglob\
is large and negative, and the cluster abundance is low. It was
easy to tell after one integration whether or not
a star is a member of the cluster.
Approximate measurements of $v_r$ were made on line;
all stars attempted turned out to be radial velocity members of \ourglob.
The four RGB stars have
a mean $v_r$ of $-176.9$ \kms, agreeing well within the errors
with the early measurement of \cite{hartwick78} of a single bright cluster
member, but somewhat larger than the value of
$-214$~\kms\ found from moderate resolution spectra
by \cite{rutledge97a}.  
Our more accurate $v_r$ for \ourglob\ should be used in
future computations
of the mass of the Galaxy which rely on the orbits of its outlying satellites.
The velocity dispersion from the four members, with no correction
for an instrumental contribution,
is $\sigma$ = 1.2$\pm$1 \kms,
reflecting the low mass of this sparse cluster;
the observed $\sigma$ is only slightly larger than the expected
instrumental uncertainties.

\subsection{Stellar Parameter Determination \label{section_teff}}

We follow the procedures developed in our earlier work on
globular cluster stars and described in
\citet{cohen01} to determine the stellar parameters for the four RGB
stars in \ourglob. \teff\ is derived by comparing reddening-corrected
broad band colors with the predictions of grids of model atmospheres.
We utilize here the grid of predicted broad band colors and
bolometric corrections of \citet{houdashelt00}
based on the MARCS stellar atmosphere code of \citet{gus75}.  

We normally utilize $V-I, V-J$ and $V-K$ colors to determine \teff.
The infrared colors were taken from 2MASS \citep{2mass1,2mass2}.
\cite{cuffey61} provided  
$P,V$ photographic photometry over a large field including the entire
cluster of \ourglob\footnote{\cite{barnes68}, who searched for variable stars in
\ourglob, defined his photometric system using Cuffey's measurements.}.
CCD photometry in $B,V$ was presented by \cite{buonanno87}, but
their field is smaller than the cluster and the calibration
of their photometry is not secure. \cite{cote91}, in the course
of a study of blue stragglers in this cluster, obtained CCD $B,V$ photometry
for a large sample of stars, but the data tables were
never published and have subsequently been lost (C\^ot\'e, private
communication, 2004).

The  two published photometric studies of \ourglob\  give
$V$ mags which are inconsistent by several tenths of a mag; 
the differences are not just a simple offset.
We therefore measured $V,I$ for our sample stars
from ANDICAM images taken for this purpose on Aug. 1, 2004
with the 1.3m telescope at CTIO operated
by the SMARTS consortium; 
these values are given in
Table~\ref{table_sample}.  ANDICAM is a dual channel cammera
constructed by the Ohio State University instrument group, but only the
optical channel was 
used\footnote{See http://www.astronomy.ohio-state.edu/ANDICAM and
http://www.astro.yale.edu/smarts.}.  Our ANDICAM program requires
photometric conditions, and additional standard star fields,
charged to our ANDICAM allocation through NOAO, are always taken for us.

We adopt a distance for \ourglob\ of 
26.2 kpc \citep{cote91} (as compared to
7.5 kpc for M13 and 10.4 kpc for M3),
with a reddening of E(B--V) = 0.036 mag from \cite{schlegel98}.
The relative extinction in various passbands is taken from
\citet{cohen81} \citep*[see also][]{schlegel98}.
The adopted stellar parameters are given in Table~\ref{table_teff}.

\subsection{Data Reduction and Analysis}

To the maximum extent possible,
the atomic data and the analysis procedures
used here are identical to those we used in our recently completed
analysis of a large sample of stars in M3 and M13 \citep{cohen05}.
In particular, see \S3 of that paper for a description of the measurement of
the equivalent widths (listed for the four stars in \ourglob\
in Table~\ref{table_eqw}), \S4 for a discussion of the atomic parameters, \S4.2 for our
adopted Solar abundances (tabulated in Table~2 of our earlier paper,
and repeated as Table~\ref{table_sun} here), 
and \S6 for a description of our abundance analysis 
procedures.  As in our earlier work,  
the abundance analysis is carried out using a current version of the LTE
spectral synthesis program MOOG \citep{sneden73}.
We employ the grid of stellar atmospheres from \cite{kurucz93} 
without convective overshoot, when available.  The template
file of suitable unblended
lines with their adopted atomic parameters is identical
to that we used in our recently completed
analysis of M3 and of M13 \citep{cohen05}.

Three of the four stars gave $v_t = 2.0$ \kms\ based on deriving
a uniform Fe abundance as a function of \eqw\ for the
large set of Fe~I lines.   The fourth star (Star~R) gave 1.8~\kms;
this value was also set to 2.0~\kms.  Lines with \eqw $>$ 175~m\AA\
were ignored, except for the Ba~II lines in the coolest star in our
sample in \ourglob.

The resulting abundance ratios for the four RGB stars in \ourglob\ are given in
Tables~\ref{table_abund_a} to \ref{table_abund_e}.
Species with only one detected line are assigned
an uncertainty of 0.10 dex.
Table~7 of \cite{cohen05} indicates
the changes in derived abundance ratios for small changes in the 
adopted stellar parameters,
the [Fe/H] for the adopted model atmosphere, or the set of \eqw\
for the lines of each species, and is appropriate for use
here as well.  
The mean abundance and 1$\sigma$ variance for the species
observed in \ourglob\ are listed in Table~\ref{table_abundsig}.

\subsection{Comments on Individual Elements \label{sec_individual}}

The oxygen abundance is derived from the forbidden lines
at 6300 and 6363~\AA.   The subtraction of the night sky emission
lines for the forbidden lines was reasonably straightforward
given that the radial
velocity of \ourglob\ is sufficiently different from 0 \kms\ 
that their \eqw\ can be reliably measured.
The C/O ratio was assumed to be Solar.
CN and Ni~I contamination is negligable \citep[see][]{cohen05}.
[O/Fe] is given with respect to Fe~II; abundance ratios for
all other elements are given with respect to Fe~I.

The deduced mean [Fe/H] of \ourglob\ value is in good agreement 
with that inferred by \cite{rutledge97b} from 
moderate dispersion spectra in the region of the infrared Ca triplet.

The Na abundance was obtained from the 5680~\AA\ doublet for all
four stars.  We have, as in our previous papers, not used any non-LTE
corrections for Na.  Calculations by \cite{gratton99} suggest
values of between 0.1 and 0.2 dex are appropriate
for our sample in \ourglob, with the coolest star having
the largest value.  Calculations by \cite{takeda03} suggest somewhat
smaller values.

The Ba abundance of star~950 in \ourglob\ appears to be
$\sim$0.2 dex  larger than that of the other cluster members
studied here.
However, the detected lines of Ba~II are all within the range where 
substantial
HFS corrections occur.  Table~7 of \cite{cohen05} shows the
very high sensitivity of the deduced Ba abundances to
small uncertainties in the equivalent width 
and microturbulence of the Ba~II lines we use.
On the other hand, the deduced [Y/Fe] is also somewhat high for this
star, while the deduced Fe is the lowest of all the stars in our sample.
Perhaps a slight adjustment of \teff\ for this star is required.
At this point, we assume this is the result of observational and
modelling uncertainties and does not indicate a real spread in
[Ba/Fe] within \ourglob, but further verification of this is desirable.

The abundances of the elements with respect to Fe, [X/Fe],
as a function of \teff\ are shown in 
Figure~\ref{figure_o_si}, covering O, Na, Mg and Si, 
Figure~\ref{figure_ca_v}, which includes Ca, Sc, Ti and V, 
Figure~\ref{figure_cr_ni}, which includes Cr, Mn,
Co and Ni, 
Figure~\ref{figure_cu_zr}, which includes Cu, Zn, Y and Zr,
and Figure~\ref{figure_ba_dy}, for Ba, La, Nd, Eu and Dy.
Note the apparent star-to-star variation in [O/Fe] and in [Na/Fe],
which becomes undetectably small, if it exists at all, for the
elements heavier than Na. The scatter for 
[Ca/Fe] and for [Ni/Fe], with 8 to 13 detected absorption lines in each
star, is remarkably small, $\le0.03$ dex over the four star sample in \ourglob.

\subsection{Abundance Spreads \label{section_spread}}

To check for the presence of
star-to-star variations in abundance ratios within the
small sample of RGB stars in \ourglob, we use
a parameter we call the ``spread ratio'' ($SR$).  The numerator of
$SR$ is the 1$\sigma$ rms variance for the sample of
four stars in \ourglob\ 
about the mean
abundance for each atomic species ($X$) with detected
absorption lines,
denoted $\sigma$; the relevant values are given in the
first  three columns  
of Table~\ref{table_abundsig}.  
The denominator of $SR$ is the total
expected uncertainty, $\sigma(tot)$, which is the sum in quadrature of 
the known contributing terms.  Included are  a term corresponding to an uncertainty
of 50~K in \teff, the same for an uncertainty of 0.2 dex in \grav,
and for an uncertainty of 0.2~\kms\ in $v_t$, and
the observed uncertainty [$\sigma(obs)$],
The parameter $\sigma(obs)$, which is calculated from data given 
in Tables~\ref{table_abund_a} to \ref{table_abund_e},
is taken as the variance about the mean abundance for a given species
in a given star, i.e. the  1$\sigma$ rms value about the mean
abundance of species $X$ in a given star/$\sqrt{N}$, where $N$ is the number of
observed lines of species $X$.  It includes contributions from
errors in the measured \eqw, random errors (i.e. between
lines of a given species) in the adopted $gf$ values, etc.
Some species, an example being Fe~I
with its very large value of $N$, 
have unrealistically small values of $\sigma(obs)$;
we adopt a minimum of 0.05 dex for this parameter.

The ratio $\sigma/\sigma(tot)$ is an indication of whether
there is any intrinsic star-to-star variation in [X/Fe].  A high value
of this ``spread ratio'', tabulated in the fifth column of
this table, suggests a high probability of intrinsic scatter for
the abundance of the species $X$. Ideally the mean $SR$ for those
elements with no star-to-star variation should be unity; for
many species the measured $SR$ is close to that value, certainly closer
here than for the sample in M3 and in M13 we studied 
earlier \citep{cohen05}.

Inspection of Table~\ref{table_abundsig} shows that for all but 
two species 
$SR < 1.0$ for the sample of four stars in \ourglob, indicating little 
sign of an intrinsic star-to-star range
in abundance.  O~I and Na~I, however, have the two largest values of
$SR$, 1.3 and 3.0 respectively.
Note that $SR$  for Mg~I is 0.5, suggesting no real 
star-to-star abundance variations for this element.
We therefore assume that the range of abundances seen in
our sample of RGB stars in \ourglob\
for Na~I and O~I represent real star-to-star abundance variations;
while no other element shows definite evidence for such variations
from this simple analysis.

\subsection{Correlated Abundance Variations of the Light Elements 
\label{section_correlated}}

C, N, O, Na, Mg, and Al are known to show correlated abundance
variations from star-to-star among the most luminous stars 
in globular clusters;
see, e.g. the review of \cite{kraft94}.
Our simple spread ratio analysis (see \S\ref{section_spread}) shows 
definite star-to-star
variations in abundance of both O and Na in our small sample of RGB stars in
\ourglob.  Variations in Mg, if present
are smaller and subtle.

It is well established that O and Na are anti-correlated among luminous
giants in globular clusters, see, e.g. \cite{kraft94}.  Furthermore,
\cite{ramirez02} compiled the data from the literature, combined it
with their own, and showed that the same linear relation can be used
to fit the O and Na data for all globular clusters studied in detail
thus far.  The latest addition to the clusters studied in detail,
NGC~2808, by \cite{carretta04b}, does so as well. 

Figure~\ref{figure_ona} shows
the relationship between Na and O abundances (both with respect to Fe)
for our sample in \ourglob.  Also superposed is the line representing 
the fit for this anti-correlation determined by \cite{sneden04} for the
luminous giants in M3, shifted by $-$0.07 dex
and +0.1 dex in the vertical and horizontal axis 
as compared to the relation we found for M13 \citep{cohen05}.
The first and last
quartiles of the O--Na anti-correlation seen by \cite{sneden04} in their
sample of luminous giants in M3 are indicated.
There is a reasonably clear anti-correlation which corresponds
well with that seen for luminous giants in other well studied
Galactic GCs \citep[see, e.g. the compilation of][]{cohen05}.
A similar correlation is detected in the outer halo GC
NGC~7006 by \cite{kraft98}.

There is marginally statistically significant evidence for a 
correlation between [Na/Fe] and [Mg/Fe] for our small sample
in \ourglob, similar to that shown in other GCs
\citep*[see, for example, figures 12 and 13 of][]{sneden04}.
The larger
uncertainty in our deduced [Mg/Fe] ratios makes this result
quite uncertain.

\section{Comparison with the Inner Halo GCs M3 and M13}

We now turn to what we can learn about the chemical history of
the Galaxy by comparing the abundance ratios in \ourglob,
at a galactocentric distance of 25 kpc, with those 
from our recent analysis of the inner halo GCs, M3 and M13 \citep{cohen05}
in M3 and M13, with $R_{GC}$ of 12 and 9 kpc respectively.
We note that the atomic parameters, the analysis procedures, and the 
software packages used are identical in both of these studies.  Hence
we should be able to detect small differences in the relative values
of the abundance ratios of these three GCs.

Table~\ref{table_abundcomp} gives
the parameter $\Delta$[X/Fe], which is, for each species
with detected lines, the mean abundance ratio
[X/Fe] for \ourglob\ with the average of the same parameter
for M3 and for M13 subtracted.  Because of the large star-to-star
differences in O/Fe seen in M3 and especially in M13, we
subtract the mean [O/Fe] for stars in M13 of luminosities comparable
to those we observe in \ourglob.  Similarly, \cite{cohen05} found a luminosity
dependence of [Mg/Fe] in M13, most probably due to the luminosity 
dependence of non-LTE corrections, which were ignored.  Again in this
case we subtract the mean [Mg/Fe] of stars of comparable luminosity
in M13 to those observed in \ourglob. 

If we allow
$\pm0.15$ dex as a tolerable range given the potential
internal and systematic errors
in the analyses of these three GCs, we find that 80\% of the
20 elements in common have a difference of 0.0$\pm0.15$ dex, with only
Si, Cr and Co\footnote{We ignore Zr as there are only a few weak
lines detected the HIRES spectrum of the coolest star in our sample 
in \ourglob.} 
outside that range.   
Figure~\ref{figure_abundcomp_m3m13}
shows the resulting differences in [X/Fe] as a function of atomic number.
The abundance ratios of [X/Fe] for Si and Co 
are each larger in \ourglob\ than they are in M3 and M13,
while that of Cr is smaller.   The largest magnitude of the set of
$\Delta$[X/Fe] occurs for Si, and is +0.23 dex.

Galactic chemical evolution produces trends in abundance ratios as a function
of metallicity.  The case of [Si/Fe] is illustrated in
Figure~\ref{figure_sife}, where metallicity is parameterized by [Fe/H];
examples for the elements Ca, Ti and Ba are given in
Figures~21, 22 and 23 of \cite{cohen05}.  
We use the same set of of high precision analyses
of GCs as in \cite{cohen05}, specifically
NGC 6528 \citep{carretta01},
NGC~6553 \citep{cohen99,carretta01}, 
47 Tuc \citep{carretta04a,james04b},
M71 \citep{ramirez02,ramirez03}, 
M5 \citep{ramirez03}, 
NGC~288 \citep{shetrone00},
NGC~362 \citep{shetrone00}, 
NGC~6752 \citep{james04a},
M3 \citep{sneden04,cohen05}, 
M13 \citep{sneden04,cohen05},
NGC~6397 \citep{thevenin01,gratton01,james04b} and
M15 \citep{sneden97}, adding in \ourglob\ as well.
To characterize the behavior of the metal-poor halo field stars, we
adopt abundance ratios from recent large surveys of 
\cite{gratton91}, \cite{mcwilliam95}, \cite{fulbright00},  and
by \cite{johnson02}.  No effort has been made to homogenize
these analyses, but since the field star surveys were carried out over
the course of more than a decade, we have corrected for the difference in the Solar
Fe abundance adopted by each.

Given that \ourglob\ has [Fe/H] 0.35 dex
smaller than the mean value for M3 and M13, we have attempted to
evaluate the correction to the difference caused by these global
trends. We can only do this for about half of the elements studied here. 
This correction is given in the last column of Table~\ref{table_abundcomp},
and is to be subtracted from the value of $\Delta$[X/Fe] to form 
$\Delta(cor)$[X/Fe].  These corrections, which do not exceed 0.1 dex
in magnitude, bring $\Delta(cor)$[Si/Fe], $\Delta(cor)$[Cr/Fe] 
and $\Delta(cor)$[Co/Fe] within the 
range consistent with no difference
(0.0$\pm0.15$ dex), while not causing any additional
elements to exceed the allowed range for equality. 
The corrections for Si, Cr and Co are shown on
Figure~\ref{figure_abundcomp_m3m13} as well.

Thus after implementing the corrections for global chemical evolution, all the
elements in common show identical abundance ratios [X/Fe] for \ourglob\ 
as for
M3 and M13, allowing for a $\pm0.15$ dex tolerance; 75\% of the 20
values of $\Delta(cor)$[X/Fe] lie within the range $-0.10$ to $+0.10$ dex.
This suggests that  the galactic
chemical evolution  of the outer halo at $R_{GC}$ 25 kpc has been
identical to that of the well studied inner halo GCs, at least up
to the time of the formation of the old globular clusters
\ourglob, M3 and M13.  In particular,
our limited evidence, based on Ba, La and Eu abundance ratios, suggests
the neutron capture processes, both $r$ and $s$, appear to have had
similar histories throughout the spatial extent of the 
halo for old GC stars.

\section{Summary}

We have carried a detailed abundance analysis for 21 elements 
in a sample of four RGB stars
in the metal poor distant outer halo globular cluster \ourglob\
([Fe/H] $-1.80$ dex).
The analyzed spectra, obtained with HIRES at the Keck Observatory,
are of high dispersion (R=$\lambda / \Delta \lambda$=35,000).
Most elements show no sign of star-to-star variation within
our limited sample.  We have, however, detected an 
anti-correlation between O and Na abundances similar to that
seen in our previous analyses of inner halo GCs as well as in
studies of relatively nearby GCs by other.  A correlation
between Mg and Na abundance may also be present.

We have compared the abundance ratios in \ourglob\ with those we
previously determined for the much closer old halo GCs M3 and 
M13 \citep{cohen05},
hoping that since all these analyses were carried out by
the same two people within a timespan of only a few months in a completely
consistent manner, with the same line lists, the same atomic
parameters, the same analysis codes and procedures, etc. that small
differences in the abundance ratios might be detectable.
We evaluate the trends of abundance ratio with metallicity
for old halo stars from our data combined with published large
surveys of halo field star abundances.  We then apply corrections
to the abundances we derived for M3 and M13 for each species,
when feasible, to extrapolate them to the 0.35 dex smaller
[Fe/H] of \ourglob.
After making such corrections,
all the elements in common show identical abundance ratios for \ourglob\ and for
M3 and M13, allowing for a $\pm0.15$ dex tolerance,
and 75\% of them are then within the tolerance $\pm$0.10 dex.
This suggests that  the galactic
chemical evolution  of the outer halo at $R_{GC}$ 25 kpc has been
identical to that of the well studied much closer inner halo GCs, at least up
to the time of the formation of the old globular clusters
\ourglob, M3 and M13.  In particular,
our limited evidence, based on Ba, La and Eu abundance ratios, suggests
the neutron capture processes, both $r$ and $s$, appear to have had
similar histories throughout the spatial extent of the 
halo for old GC stars as well.

The presence of the O/Na anti-correlation in \ourglob,
with $R_{GC}$ of 25 kpc, and the similarity of its chemical history
to that of the well studied nearby GCs, provide new constraints
on any model of GC formation in the Galactic halo.

\acknowledgements
The entire Keck/HIRES user communities owes a huge debt to 
Jerry Nelson, Gerry Smith, Steve Vogt, and many other 
people who have worked to make the Keck Telescope and HIRES  
a reality and to operate and maintain the Keck Observatory. 
We are grateful to the W. M.  Keck Foundation for the vision to fund
the construction of the W. M. Keck Observatory. 
The authors wish to extend special thanks to those of Hawaiian ancestry
on whose sacred mountain we are privileged to be guests. 
Without their generous hospitality, none of the observations presented
herein would have been possible.
This publication makes use of data from the Two Micron All-Sky Survey,
which is a joint project of the University of Massachusetts and the 
Infrared Processing and Analysis Center, funded by the 
National Aeronautics and Space Administration and the
National Science Foundation.
We are grateful to the National Science Foundation for partial support under
grant AST-0205951 to JGC.  

\clearpage


%
%
\clearpage
\begin{deluxetable}{lcrrrrrr}
\tablenum{1}
\tablewidth{0pt}
\tablecaption{The Sample of Stars in \ourglob \label{table_sample}}
\tablehead{
\colhead{ID\tablenotemark{a}} & \colhead{Coords.} & 
\colhead{V\tablenotemark{b}} &  \colhead{I\tablenotemark{b}} &
\colhead{Date Obs.} & 
\colhead{Exp. Time} &
\colhead{SNR\tablenotemark{c}} &
\colhead{$v_{r}$} \\ 
\colhead{} & \colhead{(J2000)} &
\colhead{(mag)} &   \colhead{(mag)} &
\colhead{} & 
\colhead{(sec)} &
\colhead{} & \colhead{(\kms)} }
\startdata 
H, 231 & 23 08 22.32  $-15$ 37 43 & 14.71 & 13.36 &  20/08/2003 & 800 &
    88 & $-176.8$ \\
T, 458 & 23 08 25.75  $-15$ 37 10 & 15.50 & 14.41 &  21/08/2003 & 3600 &
    92 & $-178.5$ \\
R      & 23 08 29.46  $-15$ 36 32 & 15.51 & 14.40 &  20,21/08/2003 & 3000 &
   92 & $-175.5$ \\
K, 950 & 23 08 20.83  $-15$ 36 20 & 15.77 & 14.72 & 22/08/2003 & 3600 & 
   105 & $-176.9$ \\
\enddata
\tablenotetext{a}{Alphabetical identifications are from \cite{buonanno87},
numerical ones are from \cite{cuffey61}.}
\tablenotetext{b}{Our photometry from ANDICAM images.}
\tablenotetext{c}{Signal to noise ratio in the continuum near
5865~\AA\ per 4 pixel spectral resolution element.}
\end{deluxetable}

%
%
\clearpage
\begin{deluxetable}{lccc}
\tablenum{2}
\tablewidth{0pt}
\tablecaption{Stellar Parameters\label{table_teff}}
\tablehead{
\colhead{ID\tablenotemark{a}} &
\colhead{\teff } &
\colhead{log($g$) } &
\colhead{$v_t$} \\
\colhead{} &
\colhead{(K)} &
\colhead{(dex)} &
\colhead{(km/s)} 
}
\startdata
H,231   & 4300 &  0.62 & 2.0 \\
R       & 4650 &  1.18 & 2.0  \\
T,458   & 4715 &  1.21 & 2.0 \\
K,950   & 4750 &  1.33 & 2.0  \\

\enddata
\tablenotetext{a}{Identifications as in notes to Table~\ref{table_sample}.}
\end{deluxetable}

%
%
\clearpage                                                                      
\begin{deluxetable}{llrrrrrrrrrrrrrrr}     
\tabletypesize
\scriptsize              
\tablenum{3}  
\tablewidth{0pt}                                                                
\tablecaption{Equivalent Widths\label{table_eqw}} 
\tablehead{                                                                     
\colhead{Ion} & \colhead{$\lambda$} & \colhead{$\chi_{exc}$} & 
\colhead{log $gf$} & 
\colhead{231} & \colhead{R} &  \colhead{950} & \colhead{458} \\
\colhead{} & \colhead{(\AA)} & \colhead{(eV)} & \colhead{(dex)} &
\colhead{(m\AA)} & \colhead{(m\AA)} & \colhead{(m\AA)} & \colhead{(m\AA)}
}
\startdata
  OI  &  6300.30 &  0.00 & $-$9.78 &    30.0 &    23.5 &    13.5 &    27.3 \\ 
  OI  &  6363.78 &  0.02 &$-$10.30 &     9.0 & \nodata &     8.0 &     9.5 \\ 
  NaI &  5682.63 &  2.10 & $-$0.70 &    46.0 &    32.1 &     8.0 &    16.0 \\ 
  NaI &  5688.19 &  2.10 & $-$0.42 &    61.0 &    49.6 &    15.0 &    28.5 \\ 
  NaI &  6160.75 &  2.00 & $-$1.23 &    18.5 & \nodata & \nodata & \nodata \\ 
  MgI &  4703.00 &  4.34 & $-$0.67 &   157.2 &   126.7 &   115.8 &   136.0 \\ 
  MgI &  5528.40 &  4.34 & $-$0.48 &   170.2 &   148.0 &   133.8 &   145.8 \\ 
  MgI &  5711.09 &  4.34 & $-$1.67 &    82.7 &    64.0 &    30.0 &    46.0 \\ 
  SiI &  5690.43 &  4.93 & $-$1.87 &    27.0 &    20.0 & \nodata &    23.5 \\ 
  SiI &  5948.54 &  5.08 & $-$1.23 &    42.0 &    36.0 &    33.0 &    27.0 \\ 
  SiI &  6155.13 &  5.62 & $-$0.76 &    24.0 &    20.0 &    17.0 &    31.0 \\ 
  SiI &  6237.32 &  5.62 & $-$1.01 &    15.0 & \nodata & \nodata & \nodata \\ 
  CaI &  5512.99 &  2.93 & $-$0.27 &    56.5 &    40.0 &    23.0 &    33.7 \\ 
  CaI &  5581.96 &  2.52 & $-$0.47 &    82.7 &    58.3 &    44.8 &    48.1 \\ 
  CaI &  5588.75 &  2.52 &   0.44 &   133.4 &   116.7 &    95.3 &   103.0 \\ 
  CaI &  5590.11 &  2.52 & $-$0.71 &    80.7 &    61.1 &    46.7 &    46.9 \\ 
  CaI &  5601.28 &  2.52 & $-$0.44 &    77.3 &    68.3 &    49.0 &    41.5 \\ 
  CaI &  6161.30 &  2.52 & $-$1.03 &    49.2 &    24.0 & \nodata &    22.0 \\ 
  CaI &  6162.17 &  1.90 & $-$0.09 &  \nodata  &   145.4 &   132.0 &   148.0 \\ 
  CaI &  6166.44 &  2.52 & $-$1.05 &    51.3 &    31.0 &    11.0 &    25.5 \\ 
  CaI &  6169.04 &  2.52 & $-$0.54 &    72.5 &    54.5 &    35.9 &    43.0 \\ 
  CaI &  6169.56 &  2.52 & $-$0.27 &    93.8 &    65.6 &    46.7 &    61.2 \\ 
  CaI &  6471.66 &  2.52 & $-$0.59 &    83.6 &    61.9 &    38.2 &    50.8 \\ 
  CaI &  6493.78 &  2.52 &   0.14 &   123.8 &    90.9 &    77.8 &    89.6 \\ 
 ScII &  5526.79 &  1.77 &   0.13 &    94.7 &    88.6 &    78.2 &    82.6 \\ 
 ScII &  5657.90 &  1.51 & $-$0.50 &    93.4 &    76.1 &    66.6 &    72.7 \\ 
 ScII &  5667.15 &  1.50 & $-$1.24 &    54.0 &    41.3 &    31.0 &    32.0 \\ 
 ScII &  5669.04 &  1.50 & $-$1.12 &    60.8 &    45.0 &    34.8 &    44.8 \\ 
 ScII &  5684.20 &  1.51 & $-$1.08 &    77.0 &    41.0 &    36.0 &    51.0 \\ 
 ScII &  6245.64 &  1.51 & $-$1.13 &    56.9 &    35.0 &    32.2 &    34.4 \\ 
 ScII &  6604.60 &  1.36 & $-$1.48 &    57.2 &    34.0 &    19.0 &    45.0 \\ 
  TiI &  4981.74 &  0.85 &   0.50 &   167.7 &   111.3 &    97.5 &   120.7 \\ 
  TiI &  4999.51 &  0.83 &   0.25 &   157.4 &   113.0 &    86.3 &   106.4 \\ 
  TiI &  5022.87 &  0.83 & $-$0.43 &   107.2 &    71.0 &    41.8 &    62.3 \\ 
  TiI &  5039.96 &  0.02 & $-$1.13 &   148.0 &    94.4 &    73.6 &    75.7 \\ 
  TiI &  5426.26 &  0.02 & $-$3.01 &    33.7 & \nodata & \nodata & \nodata \\ 
  TiI &  5471.20 &  1.44 & $-$1.39 &    15.0 & \nodata & \nodata & \nodata \\ 
  TiI &  5490.15 &  1.46 & $-$0.93 &    34.7 & \nodata & \nodata & \nodata \\ 
  TiI &  5644.14 &  2.27 &   0.05 &    41.0 & \nodata & \nodata & \nodata \\ 
  TiI &  5662.16 &  2.32 & $-$0.11 &    24.5 & \nodata & \nodata & \nodata \\ 
  TiI &  5937.81 &  1.07 & $-$1.89 &    16.4 & \nodata & \nodata & \nodata \\ 
  TiI &  5941.75 &  1.05 & $-$1.52 &    43.6 &    17.3 & \nodata &    14.0 \\ 
  TiI &  5953.16 &  1.89 & $-$0.33 &    41.1 & \nodata & \nodata & \nodata \\ 
  TiI &  5978.54 &  1.87 & $-$0.50 &    29.0 & \nodata & \nodata & \nodata \\ 
  TiI &  6258.10 &  1.44 & $-$0.35 &    75.0 &    37.3 &    25.4 &    35.7 \\ 
  TiI &  6261.10 &  1.43 & $-$0.48 &    72.2 &    36.5 &    18.0 &    28.7 \\ 
  TiI &  6303.76 &  1.44 & $-$1.57 &     9.0 & \nodata & \nodata & \nodata \\ 
  TiI &  6312.22 &  1.46 & $-$1.55 &     9.5 & \nodata & \nodata & \nodata \\ 
  TiI &  6743.12 &  0.90 & $-$1.63 &    45.8 & \nodata & \nodata & \nodata \\ 
 TiII &  4657.20 &  1.24 & $-$2.32 &    84.0 &    97.0 &    59.3 &    83.9 \\ 
 TiII &  4708.67 &  1.24 & $-$2.37 &    96.8 &    81.8 &    71.1 &    87.7 \\ 
 TiII &  4865.62 &  1.12 & $-$2.81 &    75.0 &    84.0 &    52.7 &    64.1 \\ 
 TiII &  5185.91 &  1.89 & $-$1.46 &    98.0 &    85.9 &    74.8 &    81.9 \\ 
 TiII &  5336.79 &  1.58 & $-$1.63 &   121.0 &   105.0 &    93.2 &   105.4 \\ 
   VI &  5670.85 &  1.08 & $-$0.43 &    35.0 & \nodata & \nodata & \nodata \\ 
   VI &  5703.57 &  1.05 & $-$0.21 &    34.3 & \nodata & \nodata & \nodata \\ 
   VI &  6081.44 &  1.05 & $-$0.58 &    29.8 & \nodata & \nodata & \nodata \\ 
   VI &  6090.22 &  1.08 & $-$0.06 &    41.7 & \nodata & \nodata & \nodata \\ 
   VI &  6199.20 &  0.29 & $-$1.28 &    32.0 & \nodata & \nodata & \nodata \\ 
   VI &  6251.82 &  0.29 & $-$1.34 &    35.4 & \nodata & \nodata & \nodata \\ 
   VI &  6274.64 &  0.27 & $-$1.67 &    21.8 & \nodata & \nodata & \nodata \\ 
   VI &  6285.14 &  0.28 & $-$1.51 &    46.2 & \nodata & \nodata & \nodata \\ 
  CrI &  5345.81 &  1.00 & $-$0.97 &   134.1 &   104.3 &    75.4 &    85.6 \\ 
  CrI &  5348.33 &  1.00 & $-$1.29 &   115.8 &    74.5 &    54.7 &    63.9 \\ 
  CrI &  5409.80 &  1.03 & $-$0.71 &   169.5 &   126.0 &    88.8 &   106.9 \\ 
  CrI &  5787.96 &  3.32 & $-$0.08 &    12.0 & \nodata & \nodata & \nodata \\ 
  MnI &  4754.04 &  2.28 & $-$0.09 &   107.2 &    74.2 &    49.8 &    62.2 \\ 
  MnI &  4783.42 &  2.30 &   0.04 &   121.2 &    95.0 &    69.2 &    72.9 \\ 
  MnI &  4823.51 &  2.32 &   0.14 &   116.5 &    75.0 &    67.5 &    78.0 \\ 
  MnI &  5537.74 &  2.19 & $-$2.02 &    28.0 & \nodata & \nodata & \nodata \\ 
  FeI &  4788.77 &  3.24 & $-$1.81 &    49.5 & \nodata & \nodata & \nodata \\ 
  FeI &  4891.50 &  2.85 & $-$0.11 &   170.7 &   144.8 &   137.7 &   142.3 \\ 
  FeI &  4919.00 &  2.86 & $-$0.34 &   168.5 &   141.0 &   117.4 &   141.4 \\ 
  FeI &  5083.34 &  0.96 & $-$2.96 &   165.2 &   131.2 &   112.3 &   126.2 \\ 
  FeI &  5166.28 &  0.00 & $-$4.20 &  \nodata  &   141.4 &   118.9 &   139.7 \\ 
  FeI &  5194.95 &  1.56 & $-$2.09 &   174.6 &   130.3 &   121.0 &   129.1 \\ 
  FeI &  5232.95 &  2.94 & $-$0.10 &   172.0 &   139.0 &   129.6 &   139.5 \\ 
  FeI &  5324.19 &  3.21 & $-$0.10 &   160.4 &   135.8 &   117.7 &   125.8 \\ 
  FeI &  5393.18 &  3.24 & $-$0.72 &   114.4 &    93.0 &    81.8 &    86.1 \\ 
  FeI &  5410.92 &  4.47 &   0.40 &    87.1 &    73.4 &    56.0 &    68.3 \\ 
  FeI &  5415.21 &  4.39 &   0.64 &   105.0 &    91.0 &    66.4 &    86.9 \\ 
  FeI &  5424.08 &  4.32 &   0.51 &   114.7 &    87.8 &    83.1 &    99.2 \\ 
  FeI &  5445.05 &  4.39 & $-$0.03 &    79.3 &    70.0 &    47.0 &    67.2 \\ 
  FeI &  5473.90 &  4.15 & $-$0.69 &    42.2 &    43.9 &    20.0 &    24.5 \\ 
  FeI &  5493.50 &  4.10 & $-$1.68 &    15.9 & \nodata & \nodata & \nodata \\ 
  FeI &  5497.52 &  1.01 & $-$2.83 &  \nodata &   147.7 &   134.3 &   143.9 \\ 
  FeI &  5501.46 &  0.96 & $-$3.05 &  \nodata &   138.6 &   118.5 &   136.5 \\ 
  FeI &  5506.79 &  0.99 & $-$2.79 &  \nodata &   146.2 &   130.0 &   142.4 \\ 
  FeI &  5525.55 &  4.23 & $-$1.08 &    21.0 & \nodata & \nodata &    13.5 \\ 
  FeI &  5554.88 &  4.55 & $-$0.35 &    33.9 &    35.0 &    22.0 &    25.6 \\ 
  FeI &  5567.39 &  2.61 & $-$2.67 &    64.8 &    39.0 & \nodata &    31.2 \\ 
  FeI &  5569.62 &  3.42 & $-$0.49 &   123.0 &    92.0 &    81.1 &    91.2 \\ 
  FeI &  5572.84 &  3.40 & $-$0.28 &   146.1 &   108.6 &   100.1 &   104.3 \\ 
  FeI &  5576.09 &  3.43 & $-$0.92 &   109.1 &    83.5 &    63.7 &    72.3 \\ 
  FeI &  5586.76 &  3.37 & $-$0.14 &   144.8 &   117.7 &   105.1 &   114.1 \\ 
  FeI &  5641.44 &  4.26 & $-$1.08 &    30.0 & \nodata & \nodata & \nodata \\ 
  FeI &  5662.52 &  4.18 & $-$0.57 &    65.0 &    48.5 &    37.2 &    39.8 \\ 
  FeI &  5679.02 &  4.65 & $-$0.82 &    25.0 & \nodata & \nodata & \nodata \\ 
  FeI &  5701.54 &  2.56 & $-$2.14 &   106.0 &    76.3 &    46.9 &    66.7 \\ 
  FeI &  5705.98 &  4.61 & $-$0.49 &    25.5 & \nodata & \nodata &    29.0 \\ 
  FeI &  5752.04 &  4.55 & $-$0.94 &    19.2 & \nodata & \nodata & \nodata \\ 
  FeI &  5753.12 &  4.26 & $-$0.69 &    52.5 &    38.9 &    30.0 &    30.4 \\ 
  FeI &  5762.99 &  4.21 & $-$0.41 &    70.2 &    44.9 &    36.5 &    48.0 \\ 
  FeI &  5775.06 &  4.22 & $-$1.30 &    22.6 & \nodata & \nodata & \nodata \\ 
  FeI &  5778.46 &  2.59 & $-$3.43 &    19.6 & \nodata & \nodata & \nodata \\ 
  FeI &  5806.72 &  4.61 & $-$0.95 &    11.0 & \nodata & \nodata & \nodata \\ 
  FeI &  5859.60 &  4.55 & $-$0.55 &    34.0 &    22.0 &    15.6 &    17.7 \\ 
  FeI &  5862.35 &  4.55 & $-$0.33 &    52.2 &    30.0 &    21.0 &    24.2 \\ 
  FeI &  5883.81 &  3.96 & $-$1.26 &    27.0 &    17.5 & \nodata &    19.0 \\ 
  FeI &  5930.17 &  4.65 & $-$0.14 &    47.4 &    36.7 &    21.3 &    18.0 \\ 
  FeI &  5934.65 &  3.93 & $-$1.07 &    42.0 &    42.0 &    29.5 &    21.8 \\ 
  FeI &  5952.72 &  3.98 & $-$1.34 &    53.3 &    35.5 &    20.0 &    24.3 \\ 
  FeI &  5956.69 &  0.86 & $-$4.50 &   110.7 &    60.7 &    34.5 &    51.2 \\ 
  FeI &  5976.79 &  3.94 & $-$1.33 &    41.7 &    18.0 &    17.5 &    19.3 \\ 
  FeI &  5983.69 &  4.55 & $-$0.66 &    37.5 & \nodata &    16.8 &    20.0 \\ 
  FeI &  6024.05 &  4.55 &   0.03 &    67.4 &    40.5 &    38.9 &    48.9 \\ 
  FeI &  6027.05 &  4.07 & $-$1.09 &    38.3 & \nodata & \nodata & \nodata \\ 
  FeI &  6055.99 &  4.73 & $-$0.37 &    29.2 &    28.0 & \nodata &    15.0 \\ 
  FeI &  6065.48 &  2.61 & $-$1.41 &   143.9 &   111.4 &    90.6 &    97.2 \\ 
  FeI &  6078.50 &  4.79 & $-$0.33 &    29.2 &    25.5 &    10.5 &    12.5 \\ 
  FeI &  6137.69 &  2.59 & $-$1.35 &   144.3 &   110.6 &    99.2 &   109.8 \\ 
  FeI &  6151.62 &  2.18 & $-$3.37 &    66.0 &    37.8 &    17.6 &    30.1 \\ 
  FeI &  6157.73 &  4.07 & $-$1.16 &    42.5 &    26.2 &     8.0 &    20.9 \\ 
  FeI &  6165.36 &  4.14 & $-$1.47 &    24.9 & \nodata & \nodata & \nodata \\ 
  FeI &  6173.34 &  2.22 & $-$2.88 &    91.0 &    62.6 &    36.0 &    51.6 \\ 
  FeI &  6180.20 &  2.73 & $-$2.65 &    55.0 &    29.8 &    14.0 &    27.7 \\ 
  FeI &  6187.99 &  3.94 & $-$1.62 &    14.0 & \nodata & \nodata & \nodata \\ 
  FeI &  6191.56 &  2.43 & $-$1.42 &   174.9 &   129.4 &    99.3 &   128.8 \\ 
  FeI &  6200.31 &  2.61 & $-$2.37 &    88.5 &    59.7 &    31.5 &    47.1 \\ 
  FeI &  6240.65 &  2.22 & $-$3.17 &    59.3 &    35.2 &    16.0 &    23.0 \\ 
  FeI &  6246.32 &  3.60 & $-$0.88 &    91.9 &    74.3 &    48.9 &    57.7 \\ 
  FeI &  6252.55 &  2.40 & $-$1.77 &   146.8 &   106.8 &   100.0 &   112.0 \\ 
  FeI &  6254.26 &  2.28 & $-$2.43 &   122.1 &    95.9 &    73.0 &    79.3 \\ 
  FeI &  6265.13 &  2.18 & $-$2.54 &   118.9 &    89.8 &    59.1 &    77.7 \\ 
  FeI &  6297.79 &  2.22 & $-$2.64 &   108.1 &    79.9 &    44.1 &    59.4 \\ 
  FeI &  6301.51 &  3.65 & $-$0.72 &    93.7 &    71.0 &    50.0 &    65.1 \\ 
  FeI &  6315.31 &  4.14 & $-$1.23 &    25.0 & \nodata & \nodata & \nodata \\ 
  FeI &  6355.03 &  2.84 & $-$2.29 &    69.3 &    45.1 &    19.0 &    33.0 \\ 
  FeI &  6380.75 &  4.19 & $-$1.38 &    22.3 & \nodata & \nodata & \nodata \\ 
  FeI &  6393.60 &  2.43 & $-$1.58 &   156.6 &   118.3 &   100.8 &   115.0 \\ 
  FeI &  6408.03 &  3.69 & $-$1.02 &    86.7 &    59.3 &    40.2 &    49.6 \\ 
  FeI &  6411.65 &  3.65 & $-$0.72 &   109.2 &    75.1 &    58.2 &    75.7 \\ 
  FeI &  6421.35 &  2.28 & $-$2.01 &   145.9 &   101.0 &    85.7 &   101.2 \\ 
  FeI &  6475.63 &  2.56 & $-$2.94 &    58.8 &    33.2 &    14.6 &    19.5 \\ 
  FeI &  6481.87 &  2.28 & $-$3.01 &    86.3 &    46.5 &    30.6 &    43.2 \\ 
  FeI &  6498.94 &  0.96 & $-$4.69 &    98.1 &    51.2 &    31.9 &    43.5 \\ 
  FeI &  6546.24 &  2.76 & $-$1.54 &   137.2 &    99.1 &    74.1 &    92.7 \\ 
  FeI &  6581.21 &  1.48 & $-$4.68 &    42.7 & \nodata & \nodata & \nodata \\ 
  FeI &  6592.91 &  2.73 & $-$1.47 &   130.9 &    97.4 &    81.7 &    90.8 \\ 
  FeI &  6593.87 &  2.43 & $-$2.37 &    99.2 &    67.4 &    48.0 &    64.2 \\ 
  FeI &  6608.02 &  2.28 & $-$3.93 &    16.0 & \nodata & \nodata & \nodata \\ 
  FeI &  6609.11 &  2.56 & $-$2.66 &    79.1 &    45.5 &    26.5 &    37.5 \\ 
  FeI &  6633.75 &  4.79 & $-$0.80 &    27.0 & \nodata & \nodata &    12.0 \\ 
  FeI &  6648.12 &  1.01 & $-$5.92 &    13.0 & \nodata & \nodata & \nodata \\ 
  FeI &  6739.52 &  1.56 & $-$4.79 &    27.0 & \nodata & \nodata & \nodata \\ 
  FeI &  6750.15 &  2.42 & $-$2.58 &   101.8 &    68.0 &    46.6 &    60.2 \\ 
  FeI &  6839.83 &  2.56 & $-$3.35 &    27.9 & \nodata & \nodata & \nodata \\ 
  FeI &  6843.65 &  4.55 & $-$0.83 &    25.6 & \nodata & \nodata & \nodata \\ 
  FeI &  6855.18 &  4.56 & $-$0.74 &    34.6 & \nodata & \nodata & \nodata \\ 
  FeI &  6861.95 &  2.42 & $-$3.85 &    20.0 & \nodata & \nodata & \nodata \\ 
  FeI &  6978.85 &  2.48 & $-$2.45 &   104.8 &    72.6 &    44.6 &    66.5 \\ 
  FeI &  6988.52 &  2.40 & $-$3.56 &    39.7 & \nodata & \nodata & \nodata \\ 
  FeI &  7022.95 &  4.19 & $-$1.15 &    38.0 & \nodata & \nodata & \nodata \\ 
  FeI &  7038.22 &  4.22 & $-$1.20 &    30.1 & \nodata & \nodata & \nodata \\ 
 FeII &  4923.93 &  2.88 & $-$1.32 &   138.4 & \nodata & \nodata & \nodata \\ 
 FeII &  5197.58 &  3.23 & $-$2.23 &    74.0 &    64.0 &    70.9 &    76.8 \\ 
 FeII &  5234.63 &  3.22 & $-$2.22 &    75.1 &    67.6 &    74.3 &    83.6 \\ 
 FeII &  5414.08 &  3.22 & $-$3.62 &    22.0 &    23.5 &    10.0 &    16.5 \\ 
 FeII &  5425.26 &  3.00 & $-$3.24 &    29.0 &    27.0 &    27.0 &    37.5 \\ 
 FeII &  5534.85 &  3.25 & $-$2.64 &    64.3 &    56.8 &    53.2 &    52.9 \\ 
 FeII &  5991.38 &  3.15 & $-$3.57 &    30.0 & \nodata &    12.0 &    27.4 \\ 
 FeII &  6149.26 &  3.89 & $-$2.69 &    18.0 &    25.2 &    16.4 &    26.9 \\ 
 FeII &  6247.56 &  3.89 & $-$2.36 &    40.0 &    31.9 &    31.0 &    31.6 \\ 
 FeII &  6369.46 &  2.89 & $-$4.20 &    10.0 & \nodata & \nodata & \nodata \\ 
 FeII &  6416.92 &  3.89 & $-$2.69 &    19.0 &    15.9 & \nodata &    15.8 \\ 
 FeII &  6516.08 &  2.89 & $-$3.45 &    64.0 &    58.3 &    47.5 &    51.8 \\ 
  CoI &  5530.79 &  1.71 & $-$2.06 &    30.0 & \nodata &     9.0 &    10.0 \\ 
  CoI &  5647.23 &  2.28 & $-$1.56 &    21.5 & \nodata & \nodata & \nodata \\ 
  CoI &  6189.00 &  1.71 & $-$2.45 &    18.0 & \nodata & \nodata & \nodata \\ 
  NiI &  5578.72 &  1.68 & $-$2.64 &    86.1 &    44.1 &    32.0 &    33.0 \\ 
  NiI &  5587.86 &  1.93 & $-$2.14 &    81.4 &    36.9 &    20.5 &    39.0 \\ 
  NiI &  5682.20 &  4.10 & $-$0.47 &    18.0 & \nodata & \nodata & \nodata \\ 
  NiI &  5748.35 &  1.68 & $-$3.26 &    41.0 & \nodata &     7.0 &    14.0 \\ 
  NiI &  5846.99 &  1.68 & $-$3.21 &    34.1 & \nodata & \nodata & \nodata \\ 
  NiI &  6128.97 &  1.68 & $-$3.33 &    30.0 &    23.5 &    10.4 &    10.5 \\ 
  NiI &  6175.37 &  4.09 & $-$0.54 &    22.0 & \nodata & \nodata & \nodata \\ 
  NiI &  6176.81 &  4.09 & $-$0.53 &    21.8 & \nodata & \nodata & \nodata \\ 
  NiI &  6177.24 &  1.83 & $-$3.51 &    15.0 & \nodata & \nodata & \nodata \\ 
  NiI &  6482.80 &  1.93 & $-$2.63 &    47.8 &    21.0 &    13.3 &    19.6 \\ 
  NiI &  6586.31 &  1.95 & $-$2.81 &    45.5 &    22.3 &     8.0 &    13.0 \\ 
  NiI &  6643.63 &  1.68 & $-$2.30 &   126.3 &    83.0 &    56.5 &    73.9 \\ 
  NiI &  6767.77 &  1.83 & $-$2.17 &   106.4 &    74.8 &    51.7 &    64.1 \\ 
  CuI &  5105.54 &  1.39 & $-$1.50 &    82.0 &    46.0 &    24.0 &    28.0 \\ 
  CuI &  5782.12 &  1.64 & $-$1.78 &    41.5 & \nodata & \nodata & \nodata \\ 
  ZnI &  4722.16 &  4.03 & $-$0.39 &    45.0 &    36.6 &    41.0 &    44.0 \\ 
  ZnI &  4810.54 &  4.08 & $-$0.17 &    46.0 &    54.6 &    52.4 &    48.5 \\ 
  YII &  4883.69 &  1.08 &   0.07 &    85.0 &    71.3 &    74.5 &    71.8 \\ 
  YII &  5087.43 &  1.08 & $-$0.17 &    71.0 &    55.0 &    56.2 &    60.1 \\ 
  YII &  5200.42 &  0.99 & $-$0.57 &    56.0 &    42.1 &    32.7 &    44.0 \\ 
  ZrI &  6127.44 &  0.15 & $-$1.06 &    17.0 & \nodata & \nodata & \nodata \\ 
  ZrI &  6134.55 &  0.00 & $-$1.28 &    13.0 & \nodata & \nodata & \nodata \\ 
  ZrI &  6143.20 &  0.07 & $-$1.10 &    10.0 & \nodata & \nodata & \nodata \\ 
 BaII &  5853.70 &  0.60 & $-$1.01 &   117.3 &    98.1 &    99.9 &   108.2 \\ 
 BaII &  6141.70 &  0.70 & $-$0.07 &   172.8 &   134.7 &   152.3 &   155.9 \\ 
 BaII &  6496.90 &  0.60 & $-$0.38 &   186.6 &   146.8 &   154.9 &   163.0 \\ 
 LaII &  5122.99 &  0.32 & $-$0.85 &    27.8 & \nodata & \nodata & \nodata \\ 
 LaII &  6390.48 &  0.32 & $-$1.41 &    21.5 & \nodata & \nodata & \nodata \\ 
 NdII &  4947.02 &  0.56 & $-$1.13 &    18.0 & \nodata & \nodata & \nodata \\ 
 NdII &  4959.12 &  0.06 & $-$0.80 &    66.6 &    34.0 &    23.5 &    32.0 \\ 
 NdII &  5092.79 &  0.38 & $-$0.61 &    44.0 &    28.5 &    17.4 &    16.9 \\ 
 NdII &  5212.35 &  0.20 & $-$0.96 &    48.0 & \nodata & \nodata & \nodata \\ 
 NdII &  5249.58 &  0.98 &   0.20 &    49.9 &    30.5 &    33.5 &    31.5 \\ 
 NdII &  5319.81 &  0.55 & $-$0.14 &    61.9 &    33.2 &    33.8 &    27.8 \\ 
 EuII &  6645.11 &  1.38 &   0.12 &    25.5 & \nodata & \nodata & \nodata \\ 
\enddata
\end{deluxetable}

\clearpage

%
%
\clearpage
\begin{deluxetable}{lclc}
\tablenum{4}
\tablewidth{0pt}
\tablecaption{Adopted Solar Abundances
\label{table_sun}}
\tablehead{
\colhead{Element} & \colhead{[X/H]\tablenotemark{a}} & 
\colhead{Element} & \colhead{[X/H]\tablenotemark{a}} \\ 
}
\startdata
O & 8.85 & Fe & 7.45 \\
Na & 6.33 & Ni & 6.25 \\
Mg & 7.54 & Cu & 4.21 \\
Al & 6.47 & Zn & 4.60 \\
Si & 7.55 & Ba & 2.13 \\
Ca & 6.36 & Y & 2.24 \\
Sc & 3.10 & Zr & 2.60 \\
Ti & 4.99 & La & 1.14 \\
V & 4.00 & Nd & 1.45 \\
Cr & 5.67 & Eu & 0.51 \\
Mn  & 5.39 & Dy & 1.10 \\
\enddata
\tablenotetext{a}{Given on a scale where log(N(H))=12.0; values in dex.}
\end{deluxetable}

\begin{deluxetable}{lcrlrrcrcrcccc}  
\tabletypesize
\footnotesize                  
\tablenum{5a}                    
\tablewidth{0pt}                
\tablecaption{Abundance Ratios: O to Mg \label{table_abund_a}} 
\tablehead{                     
\colhead{Star} &                
\colhead{[Fe/H]$_{\rm I}$}  & \colhead{$N$} &
\colhead{[Fe/H]$_{\rm II}$} & \colhead{$N$} &
\colhead{[O/Fe]} & \colhead{$N$} &
\colhead{[Na/Fe]} & \colhead{$N$} &
\colhead{[Mg/Fe]} & \colhead{$N$} \\
\colhead{} &
\colhead{$\pm\sigma/ \sqrt{N}$} & \colhead{} & 
\colhead{$\pm\sigma/ \sqrt{N}$} & \colhead{} &
\colhead{$\pm\sigma/ \sqrt{N}$} & \colhead{} & 
\colhead{$\pm\sigma/ \sqrt{N}$} & \colhead{} & 
\colhead{$\pm\sigma/ \sqrt{N}$} & \colhead{} \\
\colhead{} & 
\colhead{(dex)} & \colhead{} & \colhead{(dex)} & \colhead{} &
\colhead{(dex)} & \colhead{} & \colhead{(dex)} & \colhead{} & 
\colhead{(dex)} & \colhead{} 
}
\startdata
231 & $-$1.76 $\pm$ 0.05* & 90 & $-$1.72 $\pm$ 0.06 & 12 & 0.26 $\pm$ 0.05*& 2 &    0.10 $\pm$ 0.05  & 3 & 0.67 $\pm$ 0.16 & 3\\
R   & $-$1.77$\pm$ 0.05* & 66 & $-$1.80 $\pm$ 0.08 &  9 & 0.49 $\pm$ 0.10 & 1 &    0.19 $\pm$ 0.05* & 2 & 0.52 $\pm$ 0.06 & 3\\
458 & $-$1.79 $\pm$ 0.05* & 70 & $-$1.77 $\pm$ 0.05 & 10 & 0.55 $\pm$ 0.05*& 2 & $-$0.10 $\pm$ 0.05* & 2 & 0.55 $\pm$ 0.20 & 3\\
950 & $-$1.94 $\pm$ 0.05* & 64 & $-$1.88 $\pm$ 0.05 &  9 & 0.52 $\pm$ 0.15 & 2 & $-$0.28 $\pm$ 0.05* & 2 & 0.45 $\pm$ 0.20 & 3\\
\enddata             
\tablenotetext{*}{The minimum value of 0.05 dex has been adopted, the 
nominal calculated value is smaller.}
\end{deluxetable}    

%
\begin{deluxetable}{llrrrrrlrrrlr}  
\tabletypesize                      
\footnotesize                       
\tablenum{5b} 
\tablewidth{0pt}                
\tablecaption{Abundance Ratios: Si to V \label{table_abund_b}} 
\tablehead{                     
\colhead{Star} &                
\colhead{[Si/Fe]} & \colhead{$N$} &
\colhead{[Ca/Fe]} & \colhead{$N$} &
\colhead{[Sc/Fe]} & \colhead{$N$} &
\colhead{[Ti/Fe]} & \colhead{$N$} &
\colhead{[V/Fe]} & \colhead{$N$} \\
\colhead{} &
\colhead{$\pm\sigma/ \sqrt{N}$} & \colhead{} & 
\colhead{$\pm\sigma/ \sqrt{N}$} & \colhead{} &
\colhead{$\pm\sigma/ \sqrt{N}$} & \colhead{} & 
\colhead{$\pm\sigma/ \sqrt{N}$} & \colhead{} & 
\colhead{$\pm\sigma/ \sqrt{N}$} & \colhead{} \\
\colhead{} & 
\colhead{(dex)} & \colhead{} & \colhead{(dex)} & \colhead{} &
\colhead{(dex)} & \colhead{} & \colhead{(dex)} & \colhead{} & 
\colhead{(dex)} & \colhead{} 
}
\startdata
231 & 0.33 $\pm$ 0.09 & 4 & 0.09 $\pm$ 0.05*& 11 & 0.20 $\pm$ 0.07 & 7 & 0.16 $\pm$ 0.05*&  18 & $-$0.07 $\pm$ 0.06 & 8 \\
R   & 0.36 $\pm$ 0.10 & 3 & 0.11 $\pm$ 0.05*& 12 & 0.09 $\pm$ 0.05*& 7 & 0.12 $\pm$ 0.08 &  7 & \nodata     & 0 \\
458 & 0.45 $\pm$ 0.12 & 3 & 0.05 $\pm$ 0.05*& 12 & 0.10 $\pm$ 0.06 & 7 & 0.18 $\pm$ 0.07 &  7 & \nodata     & 0 \\
950 & 0.42 $\pm$ 0.11 & 2 & 0.07 $\pm$ 0.05 & 11 & 0.15 $\pm$ 0.05*& 7 & 0.07 $\pm$ 0.05*&  6 & \nodata     & 0 \\
\enddata     
\tablenotetext{*}{The minimum value of 0.05 dex has been adopted; the nominal 
calculated value is smaller.}                                     
\end{deluxetable}

%
\begin{deluxetable}{llrlllllrrrr}  
\tabletypesize                      
\footnotesize 
\tablenum{5c} 
\tablewidth{0pt}                
\tablecaption{Abundance Ratios: Cr to Cu \label{table_abund_c}} 
\tablehead{                     
\colhead{Star} &                
\colhead{[Cr/Fe]} & \colhead{$N$} &
\colhead{[Mn/Fe]} & \colhead{$N$} &
\colhead{[Co/Fe]} & \colhead{$N$} &
\colhead{[NiFe]} & \colhead{$N$} &
\colhead{[Cu/Fe]} & \colhead{$N$} \\
\colhead{} &
\colhead{$\pm\sigma/ \sqrt{N}$} & \colhead{} & 
\colhead{$\pm\sigma/ \sqrt{N}$} & \colhead{} &
\colhead{$\pm\sigma/ \sqrt{N}$} & \colhead{} & 
\colhead{$\pm\sigma/ \sqrt{N}$} & \colhead{} & 
\colhead{$\pm\sigma/ \sqrt{N}$} & \colhead{} \\
\colhead{} & 
\colhead{(dex)} & \colhead{} & \colhead{(dex)} & \colhead{} &
\colhead{(dex)} & \colhead{} & \colhead{(dex)} & \colhead{} & 
\colhead{(dex)} & \colhead{} 
}
\startdata
231 & $-$0.24 $\pm$ 0.09 & 4 & $-$0.24 $\pm$ 0.09 & 4  & 0.11 $\pm$ 0.05*& 3 & $-$0.06 $\pm$ 0.05*& 13 &$-$0.63 $\pm$ 0.06 & 2\\
R   & $-$0.16 $\pm$ 0.11 & 3 & $-$0.48 $\pm$ 0.13 & 3  & \nodata	  & 0 & $-$0.07 $\pm$ 0.08&  7 &$-$0.57 $\pm$ 0.10 & 1\\
458 & $-$0.29 $\pm$ 0.06 & 3 & $-$0.43 $\pm$ 0.05*& 3  & 0.18 $\pm$ 0.10 & 1 & $-$0.09 $\pm$ 0.05*&  8 &$-$0.72 $\pm$ 0.10 & 1\\
950 & $-$0.29 $\pm$ 0.05*& 3 & $-$0.39 $\pm$ 0.08 & 3  & 0.32 $\pm$ 0.10 & 1 & $-$0.09 $\pm$ 0.06&  8 &$-$0.62 $\pm$ 0.10 & 1\\
\enddata 
\tablenotetext{*}{The minimum value of 0.05 dex has been adopted; the nominal 
calculated value is smaller.}
\end{deluxetable}

%
\begin{deluxetable}{lrlllrrrrrrrrrr}  
\tabletypesize                      
\footnotesize                      
\tablenum{5d} 
\tablewidth{0pt}                
\tablecaption{Abundance Ratios: Zn to La \label{table_abund_d}} 
\tablehead{                     
\colhead{Star} &                
\colhead{[Zn/Fe]} & \colhead{$N$} &
\colhead{[Y/Fe]} & \colhead{$N$} &
\colhead{[Zr/Fe]} & \colhead{$N$} &
\colhead{[Ba/Fe]} & \colhead{$N$} &
\colhead{[La/Fe]} & \colhead{$N$} \\
\colhead{} &
\colhead{$\pm\sigma/ \sqrt{N}$} & \colhead{} & 
\colhead{$\pm\sigma/ \sqrt{N}$} & 
\colhead{} &
\colhead{$\pm\sigma/ \sqrt{N}$} & \colhead{} & 
\colhead{$\pm\sigma/ \sqrt{N}$} & \colhead{} & 
\colhead{$\pm\sigma / \sqrt{N}$} & \colhead{} \\
\colhead{} & 
\colhead{(dex)} & \colhead{} & \colhead{(dex)} & \colhead{} &
\colhead{(dex)} & \colhead{} & \colhead{(dex)} & \colhead{} & 
\colhead{(dex)} & \colhead{} 
}
\startdata
231 & $-$0.09 $\pm$ 0.08~ & 2 &  $-$0.29 $\pm$ 0.05*& 4 &  0.36 $\pm$ 0.10  & 3 & 0.24 $\pm$ 0.10 & 3 &  0.12 $\pm$ 0.16 & 2 \\
R   & $-$0.09 $\pm$ 0.08~ & 2 &  $-$0.35 $\pm$ 0.05*& 4 &	    \nodata  & 0 & 0.05 $\pm$ 0.09 & 3 &        \nodata  & 0 \\
458 & $-$0.16 $\pm$ 0.05~ & 2 &  $-$0.31 $\pm$ 0.06 & 4 &	    \nodata  & 0 & 0.34 $\pm$ 0.07 & 3 &         \nodata  & 0 \\
950 &    0.13 $\pm$ 0.05* & 2 &  $-$0.11 $\pm$ 0.08 & 4 &	    \nodata  & 0 & 0.50 $\pm$ 0.08 & 3 &         \nodata  & 0 \\
\enddata       
\tablenotetext{*}{The minimum value of 0.05 dex has been adopted; the nominal 
calculated value is smaller.}
\end{deluxetable}    

%
\begin{deluxetable}{lrrrrrrrrrrrrr}  
\tabletypesize                      
\footnotesize                      
\tablenum{5e} 
\tablewidth{0pt}                
\tablecaption{Abundance Ratios: Nd to Eu \label{table_abund_e}} 
\tablehead{                     
\colhead{Star} &                
\colhead{[Nd/Fe]} & \colhead{$N$} &
\colhead{[Eu/Fe]} & \colhead{$N$} \\
\colhead{} &
\colhead{$\pm\sigma/ \sqrt{N}$} & \colhead{} & 
\colhead{$\pm\sigma/ \sqrt{N}$} & \colhead{}  \\
\colhead{} & 
\colhead{(dex)} & \colhead{} & \colhead{(dex)} & \colhead{}  
}
\startdata
231 & 0.40  $\pm$ 0.05*  & 6 & 0.61 $\pm$ 0.10  & 1 \\
R   & 0.26  $\pm$ 0.05*  & 4 &         \nodata   & 0 \\
458 & 0.19  $\pm$ 0.07~  & 4 &         \nodata   & 0 \\
950 & 0.41  $\pm$ 0.08~  & 4 &         \nodata   & 0 \\
\enddata       
\tablenotetext{*}{The minimum value of 0.05 dex has been adopted; the nominal 
calculated value is smaller.}
\end{deluxetable}

%
%
\clearpage                                                                      
\begin{deluxetable}{lrrcrrrrrr}     
\tablenum{6}                                                                    
\tablewidth{0pt}                                                                
\tablecaption{Mean Abundances and Spread Ratios
\label{table_abundsig}}                      
\tablehead{ 
\colhead{Species} & \colhead{Mean Abund.} &
\colhead{$\sigma$} & \colhead{$\sigma$(tot)} &
\colhead{Spread Ratio\tablenotemark{a}} & \colhead{No. of} \\
\colhead{} & \colhead{[X/Fe] (dex)} &\colhead{(dex)} &
\colhead{(dex)} & \colhead{(dex)} &
\colhead{Stars\tablenotemark{e}} 
}
\startdata
OI  &   0.46 &  0.13 & 0.10 & 1.30 & 4 \\ 
NaI & $-$0.02 & 0.21 & 0.07 & 3.00 & 4 \\ 
MgI &   0.55 &  0.09 & 0.20 & 0.45 & 4 \\
SiI &   0.39 &  0.06 & 0.10 & 0.60 & 4 \\
CaI &   0.08 &  0.03 & 0.12 & 0.25 & 4 \\
ScII&   0.14 &  0.05 & 0.21 & 0.24 & 4 \\
TiI &   0.13 &  0.05 & 0.12 & 0.42 & 4 \\
TiII&   0.34 &  0.06 & 0.15 & 0.40 & 4 \\ 
VI  & $-$0.07 & \nodata& 0.23 & \nodata &  1 \\
CrI & $-$0.25 &  0.06 & 0.17 & 0.35 & 4 \\ 
MnI & $-$0.39 &  0.10 & 0.18 & 0.55 & 4 \\
FeI & $-$1.82 &  0.08 & 0.10 & 0.80 & 4 \\ 
FeII& $-$1.79 &  0.07 & 0.13 & 0.54 & 4 \\ 
CoI &   0.20 &  0.11 & 0.12 & 0.92 &  3 \\
NiI & $-$0.08 &  0.02 & 0.10 & 0.20 & 4 \\
CuI & $-$0.64 &  0.06 & 0.16 & 0.38 & 4 \\
ZnI & $-$0.03 &  0.11 & 0.13 & 0.85 & 4 \\
YII & $-$0.27 &  0.11 & 0.15 & 0.73 & 4 \\
ZrI &   0.36 & \nodata & 0.16 & \nodata  &  1 \\
BaII&   0.28 &  0.19 & 0.22 & 0.86 & 4 \\
LaII&   0.12 & \nodata & 0.18 & \nodata  &  1 \\
NdII&   0.32 &  0.11 & 0.11 & 1.00 & 4 \\
EuII&   0.61 & \nodata & \nodata & \nodata &  1 \\
\enddata
\tablenotetext{a}{This is the ratio of $\sigma$
to $\sigma$(tot).  See text.} 
\tablenotetext{e}{The number of stars in which lines of
this species were detected.}
\end{deluxetable}

%
%
\clearpage                                                                      
\begin{deluxetable}{lrrrr}     
\tablenum{7}                                                                    
\tablewidth{0pt}                                                                
\tablecaption{Comparison of Abundance Ratios in \ourglob\ to Those in M3 and M13
\label{table_abundcomp}}                      
\tablehead{ 
\colhead{Species} & \colhead{Mean Abund. NGC~7492} &
\colhead{$\sigma$} & \colhead{$\Delta$[X/Fe]} &
\colhead{[X/Fe] Corr.\tablenotemark{a}} \\
\colhead{} & \colhead{[X/Fe] (dex)} &\colhead{(dex)} &
\colhead{(NGC~7492--M3/M13) (dex)} & \colhead{(dex)} 
}
\startdata
   O &    0.46 &    0.15 &    0.02 &  \nodata \\
  Na &   $-$0.02 &    0.21 &   $-$0.10 &    0.00 \\
  Mg &    0.55 &    0.09 &    0.12 &    0.05 \\
  Si &    0.39 &    0.06 &    0.23 &   0.08 \\
  Ca &    0.08 &    0.03 &   $-$0.09 &    0.07 \\
  Sc &    0.14 &    0.05 &    0.14 &    0.07 \\
  Ti &    0.24 &    0.06 &    0.08 &    0.07 \\
  V &   $-$0.07 &   \nodata &   0.01 &  \nodata \\
  Cr &   $-$0.25 &    0.06 &   $-$0.22 &   $-$0.10 \\
  Mn &   $-$0.39 &    0.10 &   $-$0.04 &   $-$0.05 \\
  Co &    0.20 &    0.11 &    0.21 &    0.07 \\
  Ni &   $-$0.08 &    0.01 &   $-$0.01 &    0.00 \\
  Cu &   $-$0.64 &    0.06 &   $-$0.02 &  \nodata \\
  Zn &  $-0.03$ &    0.11 &    0.02 &  \nodata \\
  Y &   $-$0.27 &    0.11 &   $-$0.04 &  \nodata \\
  Zr &    0.36 &   \nodata &    0.40 &  \nodata \\
  Ba &    0.28 &    0.019&    0.05 &   $-$0.04 \\
  La &    0.12 &   \nodata &    0.03 &  \nodata \\
  Nd &    0.32 &    0.11 &    0.10 &  \nodata \\
  Eu &    0.61 &   \nodata &    0.07 &    0.00 \\
\enddata
\tablenotetext{a}{This is the approximate correction to be added to the mean 
[X/Fe] in M3/M13  to take
into account trends of [X/Fe] with [Fe/H] given the different metallicities
of the clusters, i.e. it must be subtracted from 
$\Delta$[X/Fe] to obtain $\Delta(cor)$[X/Fe].}
\end{deluxetable}

\clearpage

\begin{figure}
\epsscale{0.9}
\figurenum{2}
\plotone{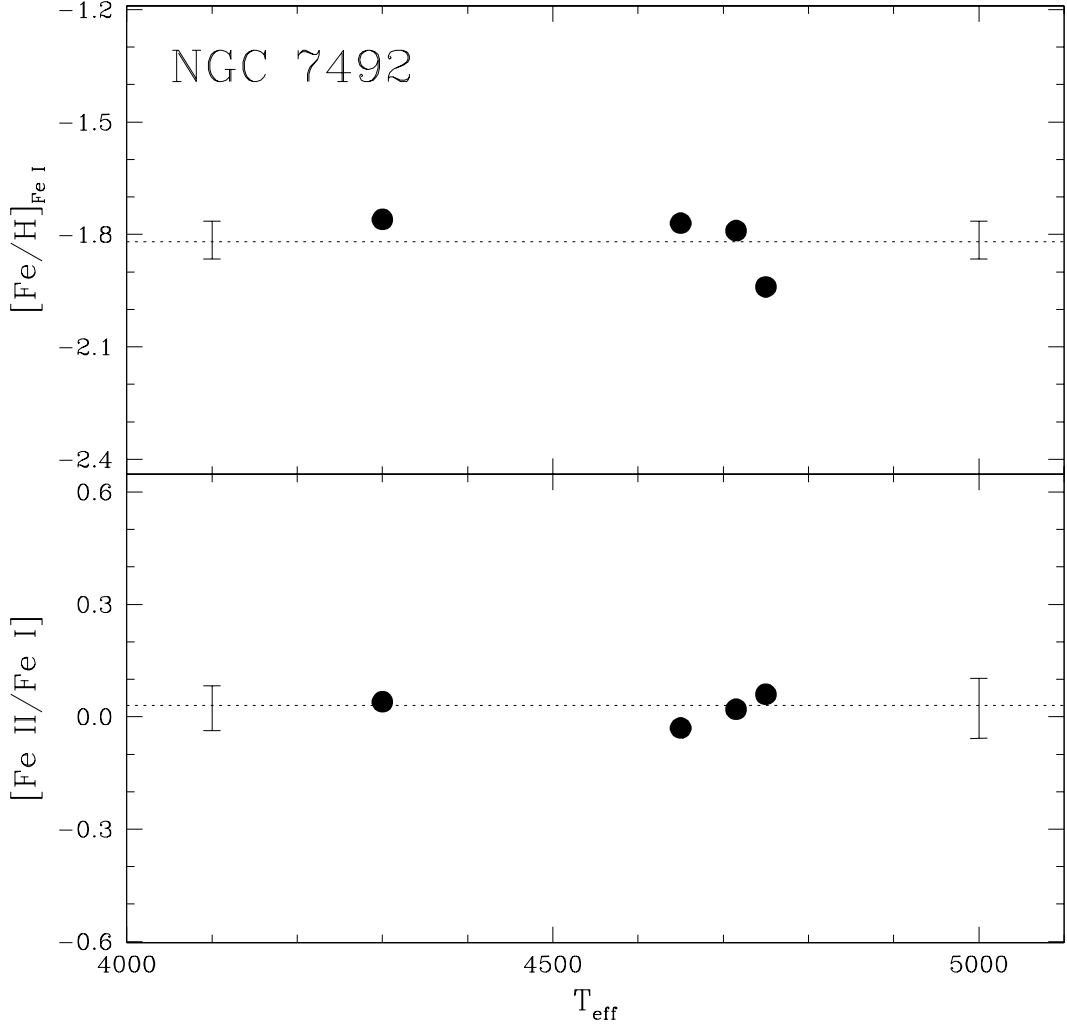}
\caption[]{The [Fe/H] from lines of Fe~I is shown as a function of \teff in
the upper panel, while the lower panel shows the ionization equilibrium of
Fe for our sample of 4 stars in \ourglob. 
The error bars on the left margin are those of the most luminous star, while 
the error bars on the right margin are those of the faintest star 
in our sample.  The dotted horizontal line indicates the mean value
for our sample in this GC.
\label{figure_feion}}
\end{figure}

\begin{figure}
\epsscale{0.9}
\figurenum{3}
\plotone{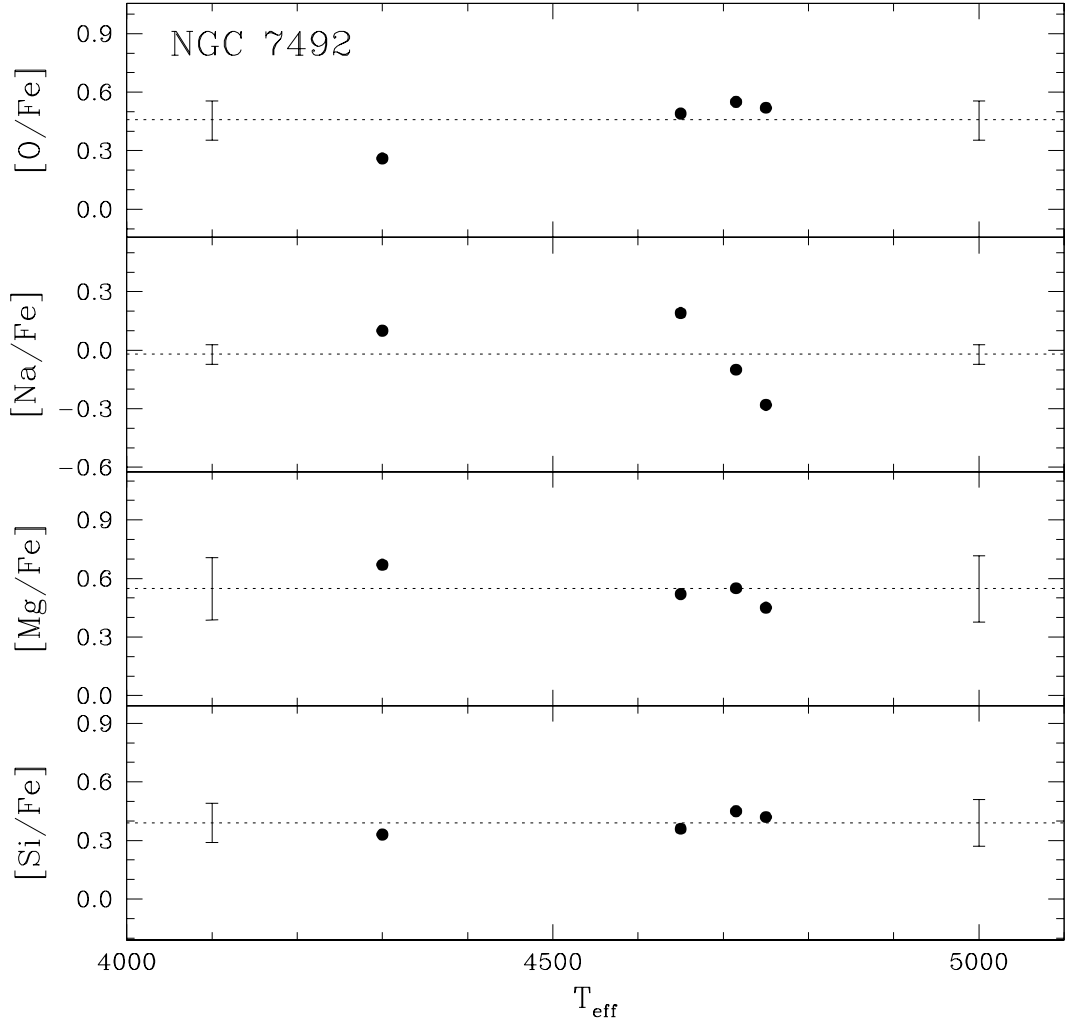}
\caption[]{[X/Fe] for the elements O, Na, Mg and Si are shown
as a function of \teff\
for our sample of 4 stars in \ourglob.
The error bars for the most luminous and least luminous stars,
as well as the cluster mean, are indicated as in Figure~\ref{figure_feion}.
\label{figure_o_si}}
\end{figure}

\begin{figure}
\epsscale{0.9}
\figurenum{4}
\plotone{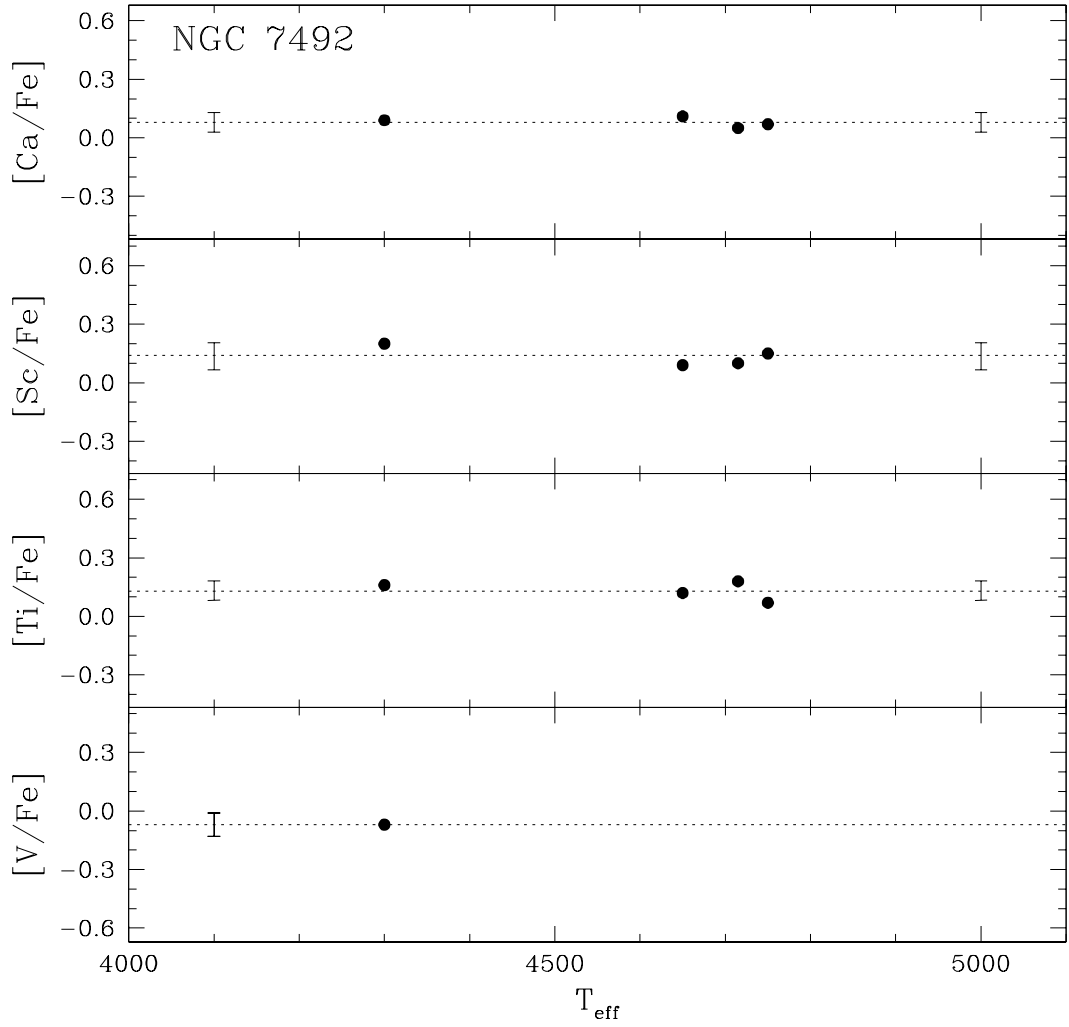}
\caption[]{Same as Figure~\ref{figure_o_si} for the elements
Ca, Sc, Ti and V in \ourglob.
\label{figure_ca_v}}
\end{figure}

\begin{figure}
\epsscale{0.9}
\figurenum{5}
\plotone{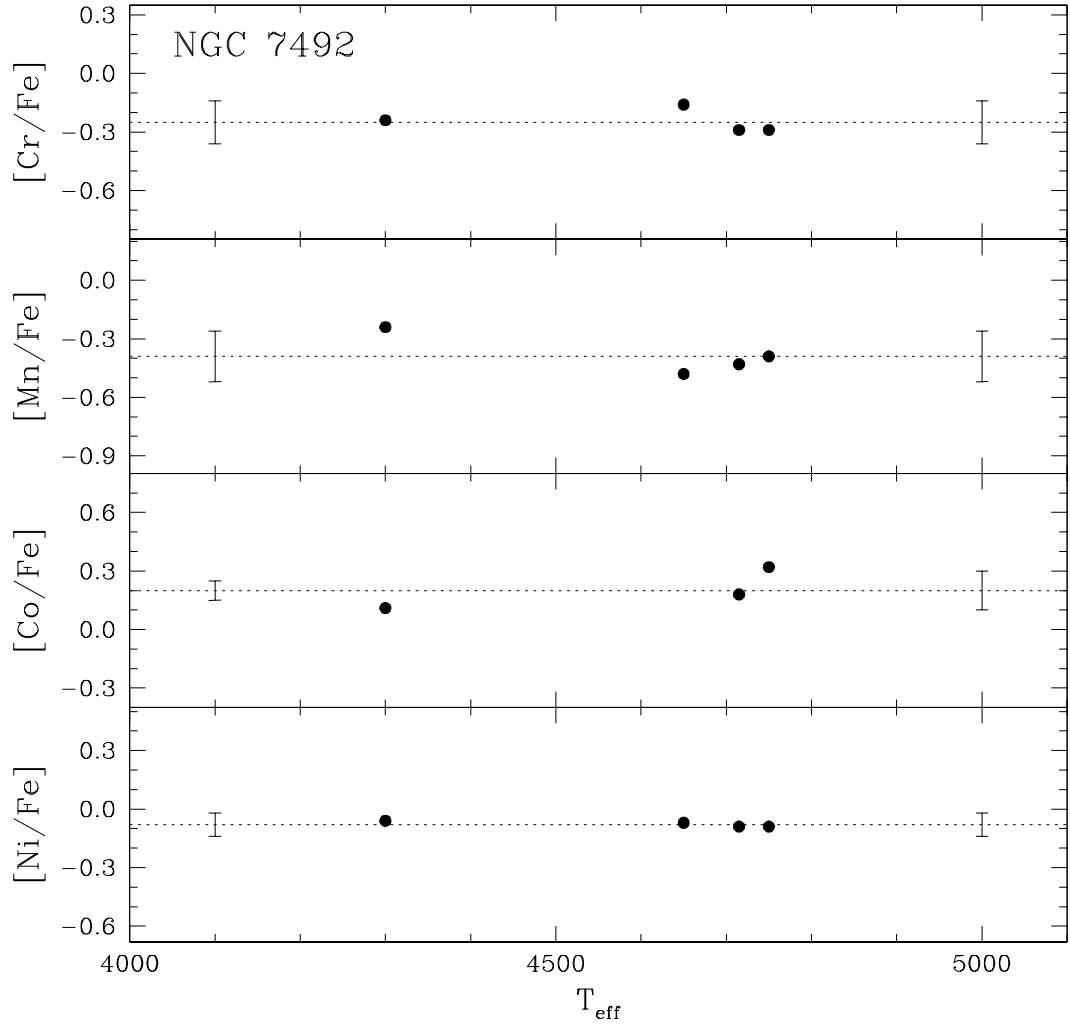}
\caption[]{Same as Figure~\ref{figure_o_si} for the elements
Cr, Mn, Co and Ni in \ourglob.
\label{figure_cr_ni}}
\end{figure}

\begin{figure}
\epsscale{0.9}
\figurenum{6}
\plotone{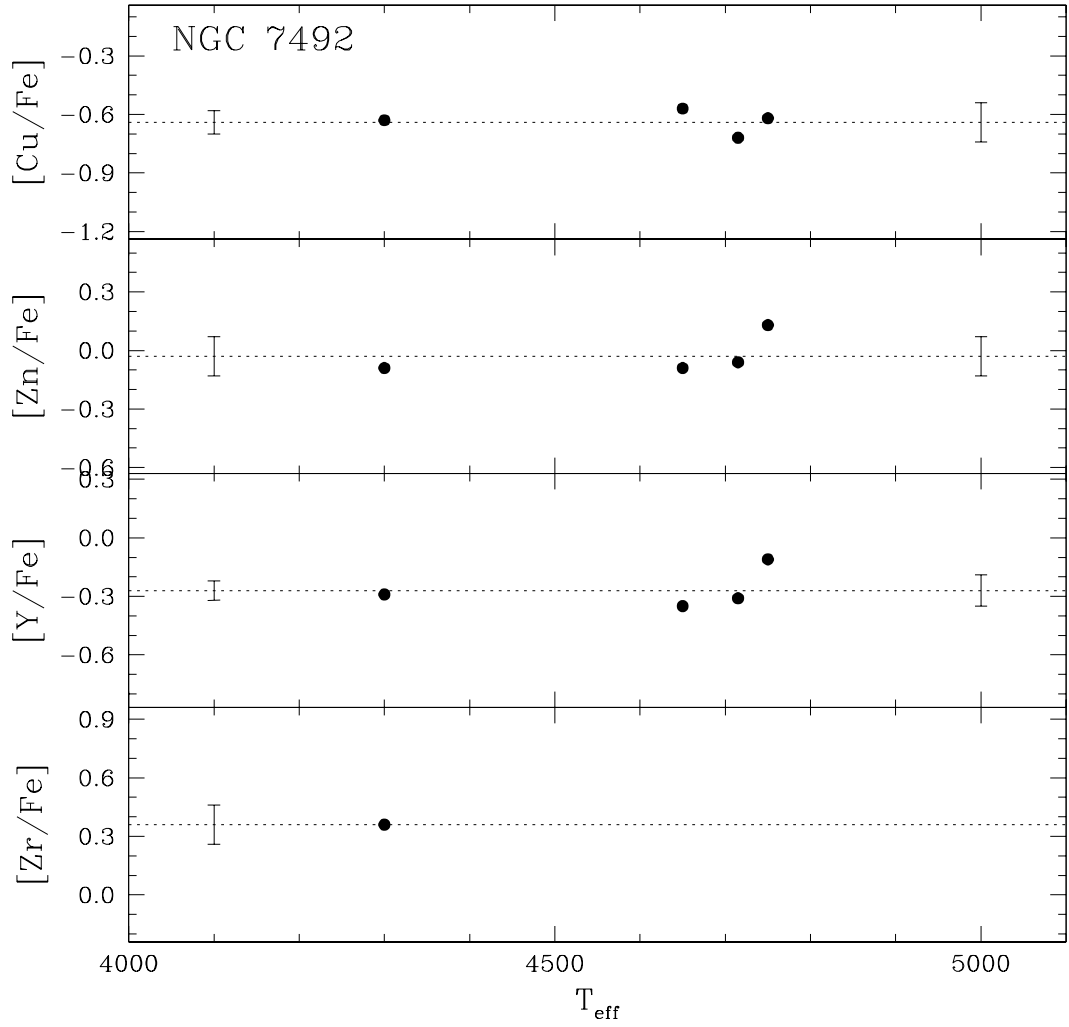}
\caption[]{Same as Figure~\ref{figure_o_si} for the elements
Cu, Zn, Y and Zr in \ourglob.
\label{figure_cu_zr}}
\end{figure}

\begin{figure}
\epsscale{0.9}
\figurenum{7}
\plotone{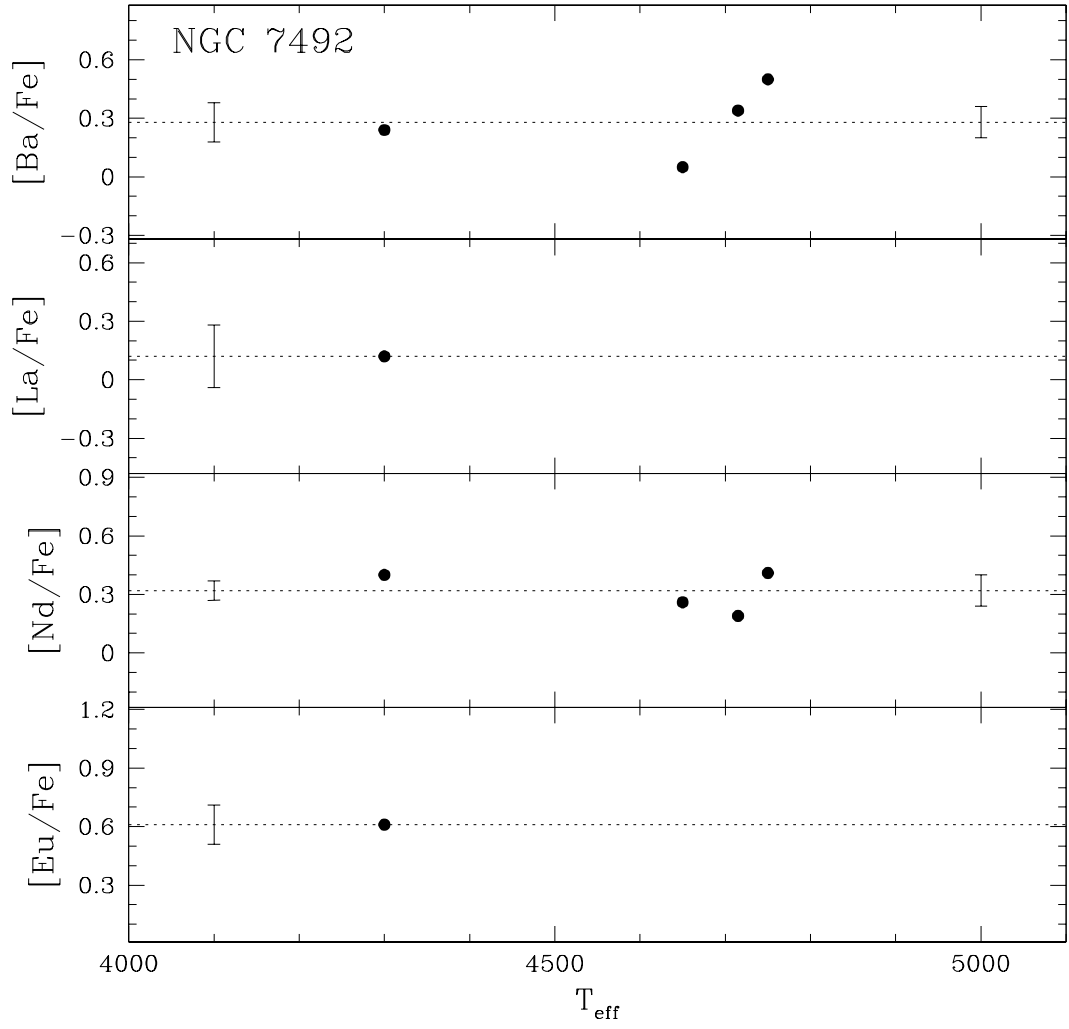}
\caption[]{Same as Figure~\ref{figure_o_si} for the elements
Ba, La, Nd and Eu in \ourglob.
\label{figure_ba_dy}}
\end{figure}

\begin{figure}
\epsscale{0.9}
\figurenum{8}
\plotone{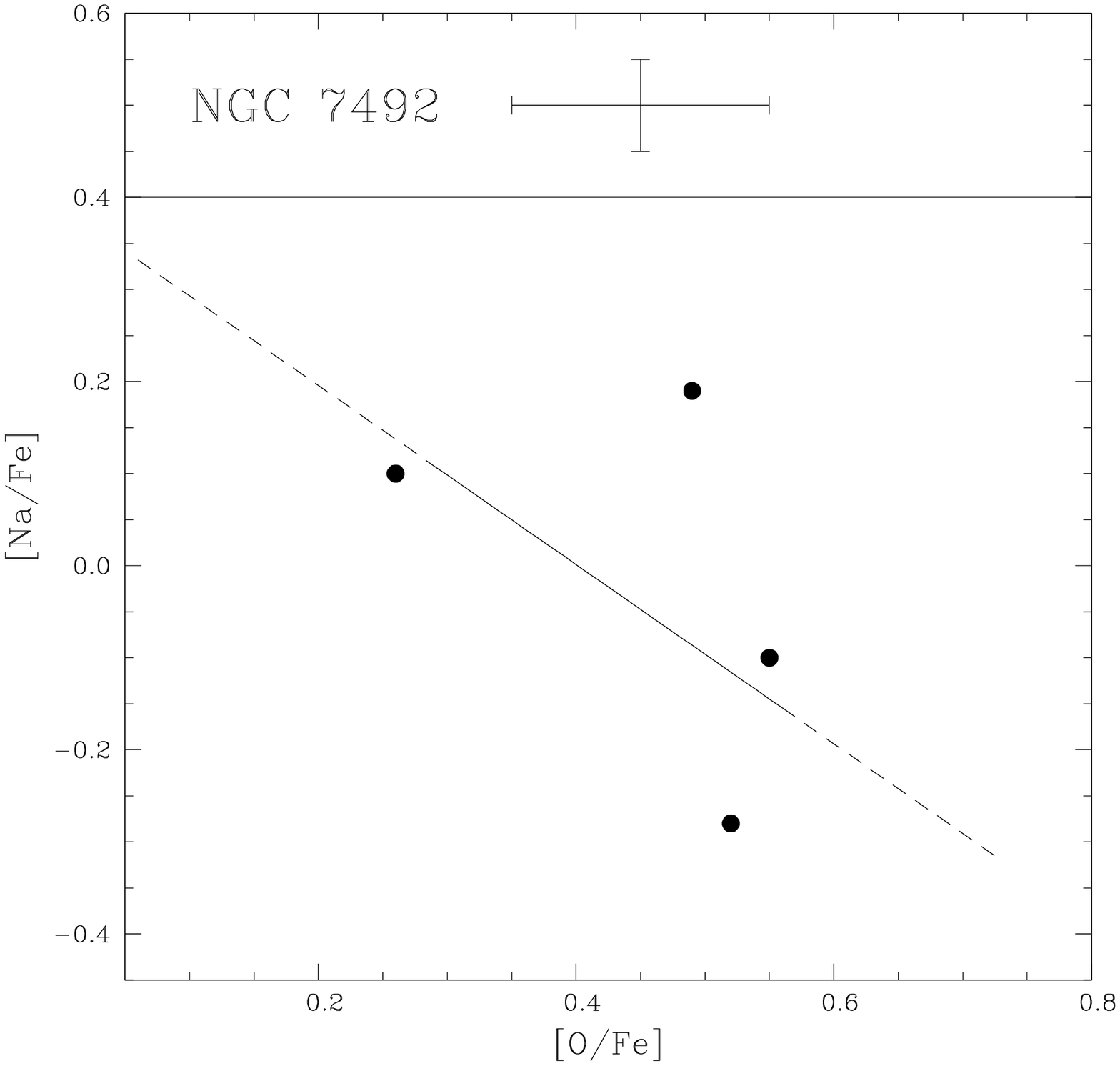}
\caption[]{The ratio [Na/Fe] is shown as a function of [O/Fe]
for our sample of four stars in \ourglob.
The error bars typical of the most luminous and least luminous stars 
in our sample are indicated.
The line represents the relationship found by \cite{sneden04},
with a shift of $-$0.07
and +0.1 dex in the vertical and horizontal axis with
respect to the relation we found for M13 \citep{cohen05};
the line is solid between the first and third quartiles of
\cite{sneden04} M3 sample and is dashed outside that regime.
\label{figure_ona}}
\end{figure}

\begin{figure}
\epsscale{0.9}
\figurenum{9}
\plotone{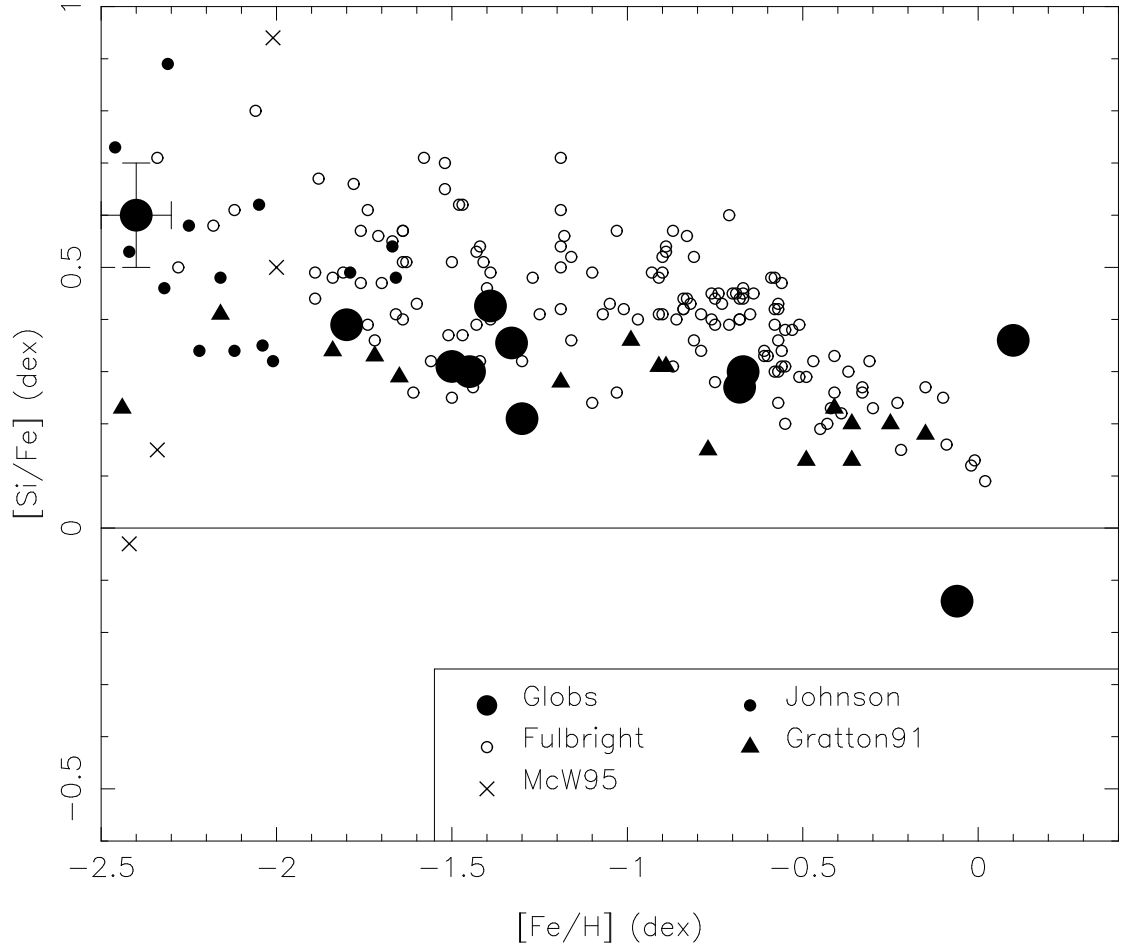}
\caption[]{The abundance ratio [Si/Fe] is shown as a function of
[Fe/H] for a sample of 13 Galactic GCs (see text for references),
indicated as large filled circles.  This
is compared to the same relationship for halo field stars 
from surveys by \cite{fulbright00}, \cite{mcwilliam95},
\cite{johnson02} and \cite{gratton91}.  An
error bar typical of the GCs is shown for the lowest metallicity GC.
\label{figure_sife}}
\end{figure}

\begin{figure}
\epsscale{0.9}
\figurenum{10}
\plotone{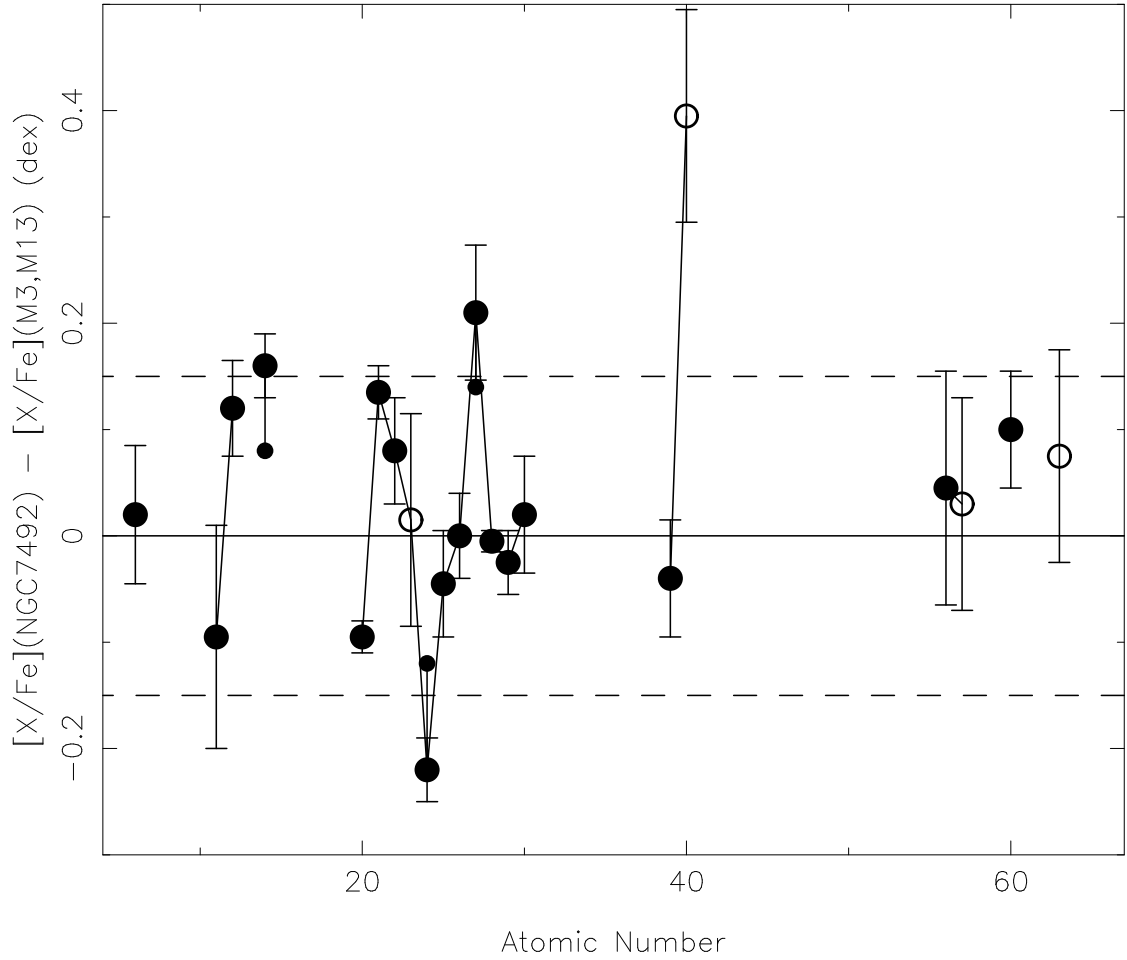}
\caption[]{The abundance ratios [X/Fe] for \ourglob, with the mean
[X/Fe] for M3 and M13 subtracted, are shown as a function of atomic number.
Open circles denote elements which have been detected in only one star 
in the \ourglob\ HIRES sample.
Lines connect the points where consecutive atomic numbers have been
detected.  The dashed horizontal lines indicate
the tolerance of $\pm0.15$ dex about equality.
Small filled circles indicate the results for Si, Cr and Co
after corrections for global trends in abundance ratio with metallicity
have been applied.
\label{figure_abundcomp_m3m13}}
\end{figure}

\end{document}